\documentclass[aps,prd,onecolumn,nofootinbib,floatfix]{revtex4-1}
\usepackage{amsmath}
\usepackage{amssymb}
\usepackage{bm}		% bold math symbols
\usepackage{graphicx}
\usepackage{url}
\usepackage{hyperref}

\newcommand{\etal}{{\it et~al.\hbox{}\/}}

\newcommand{\German}[1]{\textit{#1}}
\newcommand{\Latin}[1]{\textit{#1}}
\newcommand{\shortquote}[1]{\textit{#1}}
\newcommand{\booktitle}[1]{\textsl{#1}}
\newcommand{\journaltitle}[1]{\textsl{#1}}
\newcommand{\filmtitle}[1]{\textsl{#1}}

\newcommand{\thalf}{\tfrac{1}{2}}

\newcommand{\E}{\mathcal{E}}
\newcommand{\F}{\mathcal{F}}
\newcommand{\G}{\mathcal{G}}
\newcommand{\M}{\mathcal{M}}
\newcommand{\N}{\mathcal{N}}
\renewcommand{\O}{\mathcal{O}}
\newcommand{\R}{\mathcal{R}}
\newcommand{\U}{\mathcal{U}}
\newcommand{\V}{\mathcal{V}}
\newcommand{\boxop}{\Box}
\newcommand{\del}{\nabla}
\newcommand{\sun}{\odot}
\newcommand{\ltsim}{\lesssim}

\newcommand{\tv}{\tilde{v}}
\newcommand{\tr}{\tilde{r}}
\newcommand{\tx}{\tilde{x}}
\newcommand{\tOmega}{\tilde{\Omega}}
\newcommand{\og}{\smash{^{(0)}\!g}}	% \smash is so vertical positioning
\newcommand{\Realpart}{\text{Re}}
\newcommand{\dummy}{{\,\hbox{\rotatebox{90}{\reflectbox{\small$\sim$}}}}}
\newcommand{\adv}{\text{adv}}
\newcommand{\cons}{\text{cons}}
\newcommand{\diss}{\text{diss}}
\newcommand{\eff}{\text{eff}}
\newcommand{\far}{\text{far}}
\newcommand{\full}{\text{full}}
\newcommand{\near}{\text{near}}

\newcommand{\reg}{\text{reg}}
\newcommand{\ret}{\text{ret}}
\newcommand{\self}{\text{self}}
\newcommand{\tail}{\text{tail}}

\newcommand{\punctspace}{\hspace{0.5em}}

\begin{document}
\title{The Capra Research Program for Modelling Extreme Mass Ratio Inspirals}
\author{Jonathan Thornburg}
\affiliation{Department of Astronomy and Center for Spacetime Symmetries,
	     Indiana University, Bloomington, Indiana, USA}
%%\homepage{http://www.astro.indiana.edu/~jthorn}
\email{jthorn@astro.indiana.edu}
%
% The $ $ here lets us have the CVS-expanded  Id  keyword typeset as
% text (i.e. not in TeX math mode) -- note $$ wouldn't work (it would
% invoke TeX's display-math mode).
%
%%\date{$ $Id: capra.tex,v 1.290 2011/02/12 07:56:18 jonathan Exp $ $}

%%%%%%%%%%%%%%%%%%%%%%%%%%%%%%%%%%%%%%%%

\begin{abstract}
Suppose a small compact object (black hole or neutron star) of mass
$m$ orbits a large black hole of mass $M \gg m$.  This system emits
gravitational waves (GWs) that have a radiation-reaction effect on
the particle's motion.  EMRIs (extreme--mass-ratio inspirals) of
this type will be important GW sources for LISA.  To fully analyze
these GWs, and to detect weaker sources also present in the LISA
data stream, will require highly accurate EMRI GW templates.

In this article I outline the ``Capra'' research program to try to
model EMRIs and calculate their GWs \Latin{ab~initio}, assuming only
that $m \ll M$ and that the Einstein equations hold.  Because $m \ll M$
the timescale for the particle's orbit to shrink is too long for
a practical direct numerical integration of the Einstein equations,
and because this orbit may be deep in the large black hole's strong-field
region, a post-Newtonian approximation would be inaccurate.  Instead,
we treat the EMRI spacetime as a perturbation of the large black hole's
``background'' (Schwarzschild or Kerr) spacetime and use the methods
of black-hole perturbation theory, expanding in the small parameter
$m/M$.

The particle's motion can be described either as the result of a
radiation-reaction ``self-force'' acting in the background spacetime
or as geodesic motion in a perturbed spacetime.  Several different lines
of reasoning lead to the (same) basic $\O(m/M)$ ``MiSaTaQuWa'' equations
of motion for the particle.  In particular, the MiSaTaQuWa equations
can be derived by modelling the particle as either a point particle
or a small Schwarzschild black hole.  The latter is conceptually
elegant, but the former is technically much simpler and (surprisingly
for a nonlinear field theory such as general relativity) still yields
correct results.

Modelling the small body as a point particle, its own field is
singular along the particle worldline, so it's difficult to formulate
a meaningful ``perturbation'' theory or equations of motion there.
Detweiler and Whiting found an elegant decomposition of the
particle's metric perturbation into a singular part which is
spherically symmetric at the particle and a regular part which is
smooth (and non-symmetric) at the particle.  If we assume that the
singular part (being spherically symmetric at the particle) exerts
no force on the particle, then the MiSaTaQuWa equations follow
immediately.

The MiSaTaQuWa equations involve gradients of a (curved-spacetime)
Green function, integrated over the particle's entire past worldline.
These expressions aren't amenable to direct use in practical computations.
By carefully analysing the singularity structure of each term in a
spherical-harmonic expansion of the particle's field, Barack and Ori
found that the self-force can be written as an infinite sum of modes,
each of which can be calculated by (numerically) solving a set of
wave equations in $1{+}1$~dimensions, summing the gradients of the
resulting fields at the particle position, and then subtracting
certain analytically-calculable ``regularization parameters''.
This ``mode-sum'' regularization scheme has been the basis for much
further research including explicit numerical calculations of the
self-force in a variety of situations, initially for Schwarzschild
spacetime and more recently extending to Kerr spacetime.

Recently Barack and Golbourn developed an alternative ``$m$-mode''
regularization scheme.  This regularizes the physical metric perturbation
by subtracting from it a suitable ``puncture function'' approximation
to the Detweiler-Whiting singular field.  The residual is then
decomposed into a Fourier sum over azimuthal~($e^{im\varphi}$) modes,
and the resulting equations solved numerically in $2{+}1$~dimensions.
Vega and Detweiler have developed a related scheme that uses the
same puncture-function regularization but then solves the regularized
perturbation equation numerically in $3{+}1$~dimensions, avoiding a
mode-sum decomposition entirely.  A number of research projects are
now using these puncture-function regularization schemes, particularly
for calculations in Kerr spacetime.

Most Capra research to date has used 1st~order perturbation theory,
with the particle moving on a fixed (usually geodesic) worldline.
Much current research is devoted to generalizing this to allow the
particle worldline to be perturbed by the self-force, and to obtain
approximation schemes which remain valid over long (EMRI-inspiral)
timescales.  To obtain the very high accuracies needed to fully
exploit LISA's observations of the strongest EMRIs, 2nd~order
perturbation theory will probably also be needed; both this and
long-time approximations remain frontiers for future Capra research.
\end{abstract}

%%%%%%%%%%%%%%%%%%%%%%%%%%%%%%%%%%%%%%%%

\maketitle

%%%%%%%%%%%%%%%%%%%%%%%%%%%%%%%%%%%%%%%%%%%%%%%%%%%%%%%%%%%%%%%%%%%%%%%%%%%%%%%%
%%%%%%%%%%%%%%%%%%%%%%%%%%%%%%%%%%%%%%%%%%%%%%%%%%%%%%%%%%%%%%%%%%%%%%%%%%%%%%%%

\begin{quote}
\shortquote{This article is dedicated to the memory of Thomas Radke,
	    my late friend and colleague in many computational adventures.}
\end{quote}

%%%%%%%%%%%%%%%%%%%%%%%%%%%%%%%%%%%%%%%%%%%%%%%%%%%%%%%%%%%%%%%%%%%%%%%%%%%%%%%%
%%%%%%%%%%%%%%%%%%%%%%%%%%%%%%%%%%%%%%%%%%%%%%%%%%%%%%%%%%%%%%%%%%%%%%%%%%%%%%%%

\section{Introduction}
\label{sect-introduction}

An EMRI (extreme--mass-ratio inspiral) is a binary black hole (BH)
system (or a binary BH/neutron-star system) with a highly asymmetric
mass ratio.  That is, an EMRI consists of a small compact object (a
stellar-mass BH or neutron star) of mass $\mu M$ orbiting a large BH
of mass $M$, with the mass ratio $\mu \ll 1$.  If the small body were
a test mass ($m = 0$), then it would orbit on a geodesic of the large
BH.  However, if $m > 0$, then the system emits gravitational waves
(GWs), and there is a corresponding radiation-reaction influence on
the small body's motion.  Calculating this motion and the emitted GWs
is a long-standing research question, and is interesting both as an
abstract problem in general relativity and as an essential prerequisite
for the full success of LISA.  LISA is expected to observe GWs from
many EMRIs with $M \sim 10^6 M_\sun$ and $m \sim 10 M_\sun$
(so that $\mu \sim 10^{-5}$)~%%%
\cite{Amaro-Seoan-etal-2007:LISA-IMRI-and-EMRI-review,%%%
Gair-2009:LISA-EMRI-event-rates}.  To most effectively analyze this
LISA data -- indeed, even to \emph{detect} much weaker signals that
may also be present in the LISA data stream -- requires accurately
modelling the EMRI GWs, particularly the GW
phase~\cite{Porter-2009:LISA-data-analysis-overview}.

The small body's orbit may be highly relativistic, so post-Newtonian
methods (see, for example,~\cite[section~6.10]{Damour-in-Hawking-Israel-1987};
\cite{Blanchet-2006-living-review,Futamase-Itoh-2007:PN-review,%%%
Blanchet-2009:PN-review,Schaefer-2009:PN-review} and references therein)
may not be accurate for this problem.  Since the timescale for
radiation reaction to shrink an EMRI orbit is very long (${\sim}\, \mu^{-1} M$)
while the required resolution near the small body is very high
(${\sim}\, \mu M$), full (nonlinear) numerical-relativity methods
(see, for
example,~\cite{Pretorius-2007:2BH-review,Hannam-etal-2009:Samurai-project,%%%
Hannam-2009:2BH-review,%%%
Hannam-Hawke-2010:2BH-in-era-of-Einstein-telescope-review,%%%
Hinder-2010:2BH-review,Campanelli-etal-2010:2BH-numrel-review,%%%
Centrella-etal-2010:2BH-numrel-review}
and references therein) would be both prohibitively expensive and
insufficiently accurate for this problem.%%%
\footnote{%%%
	 The most asymmetric mass ratio yet simulated with full
	 (nonlinear) numerical relativity is $100 \,{:}\, 1$, i.e.,
	 $\mu = 10^{-2}$~\cite{Lousto-Zlochower-2011:100-to-1-mass-ratio-2BH}.
	 A number of researchers have attempted to develop
	 special methods to make EMRI numerical-relativity 
	 simulations practical, at least for systems with
	 ``intermediate'' mass ratios $\mu \sim 10^{-3}$.
	 Although promising initial results have been obtained
	 (see, for example, \cite{Bishop-etal-2003,%%%
	 Bishop-etal-2005,Sopuerta-etal-2006,Sopuerta-Laguna-2006,%%%
	 Lousto-etal-2010:intermediate-mass-2BH-numrel-Lazarus}),
	 it has not (yet) been possible to perform
	 accurate EMRI numerical evolutions lasting for
	 radiation-reaction time scales.
	 }%%%

Instead, a variety of other approximation schemes are used to model
EMRIs and their GWs.  In particular, the ``Capra'' research program,%%%
\footnote{%%%
	 The Capra research program, and the yearly Capra meetings
	 on radiation reaction in general relativity, are named
	 after the late American film director Frank Capra, famous
	 for such films as \filmtitle{It's a Wonderful Life} and
	 \filmtitle{Mr. Smith Goes to Washington} as well as the
	 World War II propaganda series \filmtitle{Why We Fight}.
	 He owned a ranch near San Diego and upon his death donated
	 part of this to Caltech.  The first Capra meeting was held
	 there in 1998.
	 }%%%
{} uses the techniques of BH perturbation theory to model the
EMRI spacetime \Latin{ab~initio} as a perturbation of the massive
central BH's Schwarzschild or Kerr spacetime, making no approximations
other than that the mass ratio $\mu \ll 1$.  In particular, the Capra
research program doesn't make any slow-motion or weak-field
approximations.

In this article I give a relatively non-technical overview of
some of the highlights of the Capra research program, focusing
on those aspects most relevant to explicitly calculating radiation-reaction
effects in various physical systems.  My goal is to give the reader
some sense of the ``flavor'' of Capra research.  The reader should
have a reasonable background in general relativity and,
for some parts of sections~\ref{sect-theory/mode-sum},
\ref{sect-theory/Detweiler-Whiting},
and~\ref{sect-theory/MiSaTaQuWa-via-modelling-as-BH},
be familiar with Green-function methods%%%
\footnote{%%%
	 We say ``Bessel function'', not ``Bessel's function'',
	 so logically the reader should be familiar with
	 ``Green-function methods'', not ``Green's-function
	 methods.%%%
	 }%%%
{} for solving linear partial differential equations (PDEs).  The
sections of this article are relatively independent and, with a few
exceptions (which should be obvious from cross-references), can be
read in any order.  In sections~\ref{sect-theory/Detweiler-Whiting}
and~\ref{sect-theory/MiSaTaQuWa-via-modelling-as-BH} I have marked
certain passages as somewhat more technical (analogous to the ``Track~2''
of Misner, Thorne, and Wheeler~\cite{MTW-1973}); this material may
be skipped if the reader so desires.

This is emphatically \emph{not} a comprehensive review -- there are
major areas of the Capra program that I only briefly mention, and
others which I omit entirely.%%%
\footnote{%%%
	 I apologise to the reader for any mistakes there
	 may be in this article, and I particularly apologise
	 to anyone whose work I've slighted or mischaracterized.
	 I welcome corrections for a future revision of this
	 article.%%%
	 }%%%
{}  Except for some of the accuracy arguments in
section~\ref{sect-accuracy}, there's no original research in this
article.  For more detailed and complete information about the Capra
research program, the reader should consult any of a number of
excellent review articles, notably those by
Poisson~\cite{Poisson-2004-living-review,Poisson-2005-GR17-plenary,%%%
Poisson-2009:self-force-review},%%%
\footnote{%%%
	 In particular, Poisson's GR17 plenary
	 lecture~\cite{Poisson-2005-GR17-plenary} contains
	 a short and relatively non-technical review of a
	 large part of the theoretical background underlying
	 the Capra research program.  I highly recommend this
	 article to the reader seeking somewhat more detail
	 than I provide in section~\ref{sect-theory}.
	 Poisson's lectures~\cite{Poisson-2009:self-force-review}
	 from the 2008 ``Mass and Motion'' summer school and
	 11th Capra meeting provides a somewhat more detailed
	 presentation of this material,
	 and his \journaltitle{Living Reviews in Relativity}
	 article~\cite{Poisson-2004-living-review}
	 gives a lengthy and detailed technical account.
	 }%%%
{} Detweiler~\cite{Detweiler-2005}, and
Barack~\cite{Barack-2009:self-force-review}.
The websites of recent Capra
meetings~\cite{Capra-2009-Bloomington-www,Capra-2010-Waterloo-www}
also include archives of meeting presentations.

A key long term goal of the Capra research program is the modelling
(and explicit calculation) of highly accurate orbital dynamics and
GW templates for generic EMRIs.  As discussed in section~\ref{sect-accuracy},
the highest-accuracy GW templates for LISA will probably require
carrying BH perturbation theory to at least 2nd~order in the
mass ratio~$\mu$, and also using special ``long-time'' approximation
schemes.  These are ambitious goals, which are still far from being
met:  most Capra research to date has been devoted to the lesser --
but still challenging -- problem of trying to model strong-field
EMRI radiation-reaction effects using 1st~order perturbation theory
and, to the best of my knowledge, no Capra GW templates have yet been
published.  I return to 2nd-order calculations in
sections~\ref{sect-accuracy} and~\ref{sect-summary}, but for the
rest of this article I consider only 1st-order calculations.

In almost all Capra calculations to date, the small body is
taken to move on a fixed geodesic worldline of the background
(Schwarzschild or Kerr) spacetime, with radiation-reaction effects
being manifest as an $\O(\mu^2)$ ``self-force'' acting on the
small body.%%%
\footnote{%%%
	 The small body's mass is $\O(\mu)$, so if it were
	 not constrained to moving on a fixed geodesic worldline,
	 the $\O(\mu^2)$ self-force would give rise to an
	 $\O(\mu)$ ``self-acceleration'' of the small body
	 away from a geodesic trajectory.%%%
	 }%%%
{}  Alternatively, we can view the small body as moving on a geodesic
of a $\O(\mu)$-perturbed spacetime.  These two perspectives can be
shown to be fully equivalent~\cite{Sago-Barack-Detweiler-2008} and
are, in some ways, analogous to Eulerian versus Lagrangian approaches
to fluid dynamics; we can use whichever is more convenient for any
given calculation.

Another important choice in self-force analyses is whether to model the
small body as a point particle or as a nonzero-sized small compact
body.  Modelling it as a nonzero-sized body is conceptually elegant
but technically difficult.  In contrast, point-particle models are
technically simpler but pose difficult conceptual and foundational
problems.  Indeed, in a nonlinear field theory such as general relativity,
the very notion of a ``point particle'' is difficult to formulate
in a self-consistent manner~\cite{Geroch-Traschen-1987}.  Remarkably,
it turns out that these difficulties can be overcome and, in fact,
point-particle models have been used for the bulk of Capra research
to date.  I discuss this point further in section~\ref{sect-theory}.

Starting from the Einstein equations, one can derive the generic
$\O(\mu)$ ``MiSaTaQuWa'' equations of motion for the small body in
an arbitrary (strong-field) curved spacetime.  These equations give
the self-force in terms of a formal Green-function integral over the
particle's entire past motion and have now been obtained in several
different ways, using both point-particle and nonzero-sized models
of the small body.

It's usually not possible to explicitly calculate the Green function
appearing in the MiSaTaQuWa equations.  Instead, practical computational
schemes are usually based on regularizing the (singular) metric-perturbation
equations for a point particle; several different ways are now known
to do this.  The regularized equations can then be solved (usually
numerically) to actually compute the self-force for a given physical
system.  Because these calculations are in many cases both conceptually
difficult and computationally demanding, new techniques are often
first developed on simpler electromagnetic or scalar-field ``model''
systems.  These retain many of the basic conceptual features of the
gravitational case while greatly simplifying the gauge choice%%%
\footnote{%%%
	 As discussed by Barack and Ori~\cite{Barack-Ori-2001},
	 the self-force is highly gauge-dependent in a somewhat
	 unobvious non-tensorial manner.  (For example, there exists
	 a gauge in which the self-force vanishes.  Essentially,
	 the gauge transformation follows the small body as it
	 spirals in to the massive BH.)  There are thus considerable
	 benefits to computing gauge-invariant effects, an approach
	 particularly championed by Detweiler.%%%
	 }%%%
{} and the resulting computations.

We can categorize Capra self-force calculations along several
dimensions of complexity:
\begin{itemize}
\item	The background spacetime may be either Schwarzschild or Kerr.
\item	The field equations may be for the scalar-field,
	electromagnetic, or the full gravitational case.
\item	The small body may be stationary, in an equatorial circular
	orbit, in a generic (non-circular) equatorial orbit, or in
	a fully generic (inclined non-circular) orbit in Kerr spacetime.
\end{itemize}

The outline of the remainder of this article is as follows:
In section~\ref{sect-theory} I discuss some of the key
theoretical foundations of the Capra program including 
the Barack-Ori mode-sum regularization,
the Detweiler-Whiting decomposition of a point particle's metric perturbation,
several different derivations of the basic 1st-order ``MiSaTaQuWa''
equations of motion for a small compact body moving in a curved spacetime, 
the Barack-Golbourn and Vega-Detweiler puncture-function regularizations
and the self-force computational schemes derived from them,
and the decomposition of self-force effects into
conservative and dissipative parts.
In section~\ref{sect-Barack-Sago} I summarize a recent self-force
calculation of Barack and Sago~\cite{Barack-Sago-2010}, which provides
an almost complete solution of the 1st-order self-force problem for
a particle moving on a fixed geodesic orbit in Schwarzschild spacetime.
In section~\ref{sect-accuracy} I roughly estimate LISA's accuracy
requirements for GW templates, and outline some of the issues in
trying to model EMRI orbital dynamics for long (orbital-decay) times
to construct such templates.
Finally, in section~\ref{sect-summary} I summarize the progress
of the Capra program to date and discuss some of its likely future
prospects.

Throughout this article I use $c = G = 1$ units
and a $(-,+,+,+)$ metric signature.
I use the Penrose abstract-index notation,
with $abcde$ as spacetime indices.
$\delta(\cdot)$ is the Dirac $\delta$-function,
$\tau$ denotes proper time along the small body's worldline,
and a subscript $p$ denotes evaluation at the small body (particle)'s
current position.
$\mu \ll 1$ is the EMRI system's mass ratio
and $M$ the central BH's mass.
Apart from these, the notation in this article varies somewhat from
section to section; it's always described at the start of each section.

%%%%%%%%%%%%%%%%%%%%%%%%%%%%%%%%%%%%%%%%%%%%%%%%%%%%%%%%%%%%%%%%%%%%%%%%%%%%%%%%

\section{Theoretical Background}
\label{sect-theory}

In this section I discuss some of the main theoretical background
and formalisms which underlie the Capra research program.%%%
\footnote{%%%
	 My exposition in parts of this section draws
	 heavily on that of Poisson's GR17 plenary
	 lecture~\cite{Poisson-2005-GR17-plenary}.%%%
	 }%%%

A key early result of Capra research was the derivation in several
different ways of the basic 1st-order equations of motion for a
small compact body moving in a strong-field curved spacetime.
These equations were first derived in 1997 by
Mino, Sasaki, and Tanaka~\cite{Mino-Sasaki-Tanaka-1997} and
Quinn and Wald~\cite{Quinn-Wald-1997}, and (abbreviating the
authors' names) are now known as the ``MiSaTaQuWa'' equations.

The MiSaTaQuWa equations involve gradients of a curved-spacetime
Green function, integrated over the particle's entire past worldline.
We can rarely calculate the Green function explicitly, so the
MiSaTaQuWa equations aren't useful for practical computations.
In section~\ref{sect-theory/mode-sum} I discuss
the ``mode-sum regularization'' computational scheme due originally
to Barack and Ori~\cite{Barack-Ori-2000}.  This scheme regularizes
each mode of a spherical-harmonic decomposition of the (singular)
scalar-field or metric perturbation, then solves numerically for
each regularized mode in $1{+}1$~dimensions.  This scheme has been
the basis for much further research, including many practical
self-force calculations.

In 2003 Detweiler and Whiting~\cite{Detweiler-Whiting-2003} found
a Green-function decomposition -- and a corresponding decomposition
of the metric perturbation due to a small particle -- into singular
and radiative fields, which greatly aids understanding the self-force
and related phenomena.  In section~\ref{sect-theory/Detweiler-Whiting}
I discuss this decomposition and the related Detweiler-Whiting
``postulate'' concerning the physical significance of the singular
and radiative fields.

In section~\ref{sect-theory/MiSaTaQuWa-via-modelling-as-point-particle}
I outline how the Detweiler-Whiting postulate allows the MiSaTaQuWa
equations to be derived via modelling the small body as a point particle.
This derivation of the MiSaTaQuWa equations is quite simple, but it
does involve the introduction of point particles and the assumption
of the Detweiler-Whiting postulate.

As an alternative, in
section~\ref{sect-theory/MiSaTaQuWa-via-modelling-as-BH} I outline
a different derivation of the MiSaTaQuWa equations, this time modelling
the small body as a small nonrotating (Schwarzschild) BH.  This
derivation is technically more difficult than the point-particle
derivation, but it avoids both the introduction of point particles
and the assumption of the Detweiler-Whiting postulate.

Recently researchers have developed several new analyses of the
self-force problem, leading to much more satisfactory derivations
of the MiSaTaQuWa (and analogous) equations.  These new analyses are
fully rigorous and resolve a number of past conceptual difficulties
as well as opening promising avenues for further research.
I (very) briefly outline these analyses in
section~\ref{sect-theory/MiSaTaQuWa-other-derivations}.

Recently two groups have developed alternate ``puncture-function''
regularization schemes for self-force computations.  Both schemes
first subtract from the physical field a suitable ``puncture function''
approximation to the Detweiler-Whiting singular field, leaving a
regular remainder field.
Barack, Golbourn, and their coauthors~\cite{Barack-Golbourn-2007,%%%
Barack-Golbourn-Sago-2007,Dolan-Barack-2011}
developed an ``$m$-mode'' regularization scheme which then decomposes
the perturbation equation into a Fourier sum over azimuthal~($e^{im\varphi}$)
modes, and finally solves numerically for each mode in $2{+}1$~dimensions.
Vega, Detweiler, and their
coauthors~\cite{Vega-Detweiler-2008:self-force-regularization,%%%
Vega-etal-2009:self-force-3+1-primer} have developed a different
puncture-function regularization scheme that numerically solves the
regularized equation directly in $3{+}1$~dimensions, avoiding a
mode-sum decomposition entirely.  I describe both of these schemes
in section~\ref{sect-theory/puncture-fn}.

The self force can be decomposed into conservative (time-symmetric)
and dissipative (time-antisymmetric) parts.
In section~\ref{sect-theory/conservative-vs-dissipative}
I discuss this decomposition and its physical significance.

%%%%%%%%%%%%%%%%%%%%%%%%%%%%%%%%%%%%%%%%

\subsection{The Barack-Ori Mode-Sum Regularization}
\label{sect-theory/mode-sum}

The MiSaTaQuWa equations (discussed further in
sections~\ref{sect-theory/Detweiler-Whiting}--\ref{sect-theory/MiSaTaQuWa-other-derivations})
give the self-force in terms of the gradient of a curved-spacetime
Green function, integrated over the entire past history of the small
body.  (The integral must be cut off infinitesimally before the small
body's current position.)  For most physically-interesting systems
we can't explicitly calculate the Green function, so the MiSaTaQuWa
equations aren't useful for practical calculations.  Instead, almost
all practical self-force calculations use other regularized
reformulations of the (singular) scalar-field, electromagnetic, or
metric-perturbation equations.

Building on earlier suggestions of Ori~\cite{Ori-1995,Ori-1997},
in 2000 Barack and Ori~\cite{Barack-Ori-2000} proposed a practical
``mode-sum'' regularization of the field equations, initially for
the model problem of a scalar particle moving in Schwarzschild
spacetime.  Barack and various coauthors~\cite{Barack-2000,Barack-2001,%%%
Barack-etal-2002,Barack-Ori-2002,Barack-Lousto-2002,Barack-Ori-2003}
soon extended this to include electromagnetic and gravitational
particles in a somewhat wider class of spherically symmetric
black hole spacetimes, as well as scalar-field, electromagnetic,
and gravitational particles in Kerr spacetime~\cite{Barack-Ori-2003:Kerr}.
The mode-sum regularization has been the basis for much further
research including many practical self-force calculations.

To explain the mode-sum regularization, I consider the scalar-field
case -- this contains the essential ideas, but is technically much
simpler than the full gravitational case.
Thus, consider a point particle of scalar charge~$q$, moving along
a timelike worldline~$\Gamma = \Gamma(\tau)$ in a background
Schwarzschild spacetime with metric $g_{ab}$ and covariant derivative
operator~$\del_a$.  In this section we raise and lower all indices
with the background Schwarzschild metric $g_{ab}$, and we take
$(t,r,\theta,\varphi)$ to be the usual Schwarzschild coordinates.

We take the scalar field $\phi$ to satisfy the usual scalar wave
equation
\begin{equation}
\boxop \phi
	= -4 \pi q \! \int_{-\infty}^{+\infty} \!
		      \frac{\delta^4\bigl(x^a - \Gamma^a(\tau')\bigr)}
			   {\sqrt{-g}}
		      \, d\tau'
	=: S
							    \text{\punctspace ,}
					      \label{eqn-theory/scalar-wave-eqn}
\end{equation}
where the integral extends over the entire worldline of the particle,
and we define $S$ to be the source term (right hand side).
We assume that if the scalar field were regular (non-singular) at the
particle, it would exert a force
\begin{equation}
F_a = q \, (\del_a \phi)_p
				       \label{eqn-theory/scalar-field-force-law}
\end{equation}
on the particle.

The scalar wave equation~\eqref{eqn-theory/scalar-wave-eqn}
can be formally solved by means of a retarded Green function~$\G(x,x')$,
\begin{equation}
\phi(x) = \int_{-\infty}^{+\infty} \! \G\bigl(x,\Gamma(\tau')\bigr) \, d\tau'
							    \text{\punctspace ,}
			    \label{eqn-theory/mode-sum/phi-formal-Green-fn-soln}
\end{equation}
where the Green function~$\G(x,x')$ satisfies
\begin{equation}
\boxop \G(x,x') = - \frac{4\pi}{\sqrt{-g}} \, \delta^4(x^a - {x'}^a)
\end{equation}
and incorporates the appropriate causality relationships.

Ignoring some terms which aren't relevant here, the self-force on
the particle at the worldline event~$x = (t,r,\theta,\varphi)$ can
then be shown to be given by
\begin{equation}
F_a(x) = q^2 \! \int_{-\infty}^{x^-} \!
		\del_a \G\bigl(x;\Gamma(\tau')\bigr) \, d\tau'
							    \text{\punctspace ,}
				  \label{eqn-theory/mode-sum/F=int-del-Green-fn}
\end{equation}
where the upper limit $x^-$ means that the integral extends over the
entire past worldline of the particle prior to (but not including)
the event~$x$.

Now consider a spherical-harmonic decomposition of the
scalar field~$\phi$ and the self-force~$F_a$,
\begin{subequations}
				   \label{eqn-theory/mode-sum/Ylm-decomposition}
\begin{eqnarray}
\phi(x)
	& = &	\sum_{\ell=0}^\infty
		\sum_{m=-\ell}^{+\ell} \!
		Y_{\ell m}(\theta,\varphi) \phi^{\ell m}(t,r)
									\\
F_a(x)
	& = &	\sum_{\ell=0}^\infty
		\sum_{m=-\ell}^{+\ell} \!
		Y_{\ell m}(\theta,\varphi) F_a^{\ell m}(t,r)
							    \text{\punctspace ,}
									%%%\\
\end{eqnarray}
\end{subequations}
and sum over the azimuthal mode number~$m$ by defining
\begin{equation}
F_a^\ell(t,r,\theta,\varphi)
	= \sum_{m=-\ell}^{+\ell} \!
	  Y_{\ell m}(\theta,\varphi) F_a^{\ell m}(t,r)
							    \text{\punctspace ,}
					       \label{eqn-theory/mode-sum/F-ell}
\end{equation}
so that the self-force is given by
\begin{equation}
F_a = \sum_{\ell=0}^\infty F_a^\ell
					 \label{eqn-theory/mode-sum/F=sum-F-ell}
							    \text{\punctspace .}
\end{equation}

It turns out that each individual spherical-harmonic mode~$F_a^\ell$
is finite, but the sum over~$\ell$ of these modes
in~\eqref{eqn-theory/mode-sum/F=sum-F-ell} diverges.  However,
by carefully analysing the divergence of the scalar field and its
Green function near the particle, Barack and Ori showed that the
related sum
\begin{equation}
\sum_{\ell=0}^\infty
\left( F_a^\ell - [A_a L + B_a + C_a/L] \right)
							    \text{\punctspace ,}
\end{equation}
(where $L = \ell+\thalf$, and where the quantities $A_a$, $B_a$,
and $C_a$ are described below) \emph{does} in fact converge and
moreover, that the self-force is given by
\begin{equation}
F_a = \sum_{\ell=0}^\infty
      \left( F_a^\ell - [A_a L + B_a + C_a/L] \right) - Z_a
				   \label{eqn-theory/mode-sum/F=sum-regularized}
							    \text{\punctspace ,}
\end{equation}
where the ``regularization parameters'' $A_a$, $B_a$, $C_a$, and $Z_a$
are independent of $\ell$ and can be calculated semi-analytically as
elliptic integrals depending on the worldline~$\Gamma$ and the worldline
event~$x$.

[More generally, additional even-power terms $D^{(2)}_a\!/L^2$,
$D^{(4)}_a\!/L^4$, $D^{(6)}_a\!/L^6$, \dots, can be added to the
$A_a L + B_a + C_a/L$ inner sum in the mode-sum self-force
formula~\eqref{eqn-theory/mode-sum/F=sum-regularized} (with
corresponding adjustments to the definition of $Z_a$).  As discussed by
Detweiler, Messaritaki, and Whiting~\cite{Detweiler-Messaritaki-Whiting-2003},
if the additional regularization parameters $D^{(k)}_a$ can be
explicitly calculated (analytically or semi-analytically), then
adding these extra terms can greatly accelerate the convergence of
the sum over~$\ell$.  I discuss the numerical treatment of this
infinite sum below.]

By virtue of the spherical symmetry of the Schwarzschild background,
the spherical-harmonic
decomposition~\eqref{eqn-theory/mode-sum/Ylm-decomposition} separates 
the scalar wave equation~\eqref{eqn-theory/scalar-wave-eqn}.
That is, each individual $\phi^{\ell m}(t,r)$ now satisfies a
linear wave equation in $1{+}1$~dimensions on the Schwarzschild
background,
\begin{equation}
\boxop \phi^{\ell m} + V_\ell(r) \phi^{\ell m}
	= S_{\ell m}(t) \, \delta\bigl(r - r_p(t)\bigr)
							    \text{\punctspace ,}
					  \label{eqn-theory/mode-sum/phi-lm-eqn}
\end{equation}
where the potential $V_\ell(r)$ and source term $S_{\ell m}(t)$ are
known analytically, and where $r_p(t)$ is the particle's (known)
Schwarzschild radial coordinate (which is time-dependent if the
worldline is anything other than a circular geodesic orbit).
The wave equation~\eqref{eqn-theory/mode-sum/phi-lm-eqn}
can then be solved numerically in the time domain to find the
field~$\phi^{\ell m}$.

Finally, each individual self-force mode~$F_a^{\ell m}$ can be
calculated from the $\ell m$ component of the basic
force law~\eqref{eqn-theory/scalar-field-force-law},
\begin{equation}
F_a^{\ell m} = q \left(\del_a \phi^{\ell m}\right)_p
							    \text{\punctspace ,}
\end{equation}
and then $F_a^\ell$ can be calculated from~\eqref{eqn-theory/mode-sum/F-ell}.

Alternatively, we can take a frequency-domain approach, augmenting the
spherical-harmonic expansions~\eqref{eqn-theory/mode-sum/Ylm-decomposition}
with a Fourier transform in time, i.e., we can replace those expansions
with
\begin{widetext}
\begin{subequations}
\begin{eqnarray}
\phi(x)
	& = &	\int\limits_\omega
		\sum_{\ell=0}^\infty
		\sum_{m=-\ell}^{+\ell} \!
		Y_{\ell m}(\theta,\varphi) e^{i\omega_m t}
					   \phi^{\omega\ell m}(r)
		\, d\omega
									\\
F_a(x)
	& = &	\int\limits_\omega
		\sum_{\ell=0}^\infty
		\sum_{m=-\ell}^{+\ell} \!
		Y_{\ell m}(\theta,\varphi) e^{i\omega_m t}
					   F_a^{\omega\ell m}(r)
		\, d\omega
							    \text{\punctspace ,}
									%%%\\
\end{eqnarray}
and correspondingly replace the $m$~summation~\eqref{eqn-theory/mode-sum/F-ell}
with
\begin{equation}
F_a^\ell(t,r,\theta,\varphi)
	= \int\limits_\omega
	  \sum_{m=-\ell}^{+\ell} \!
	  Y_{\ell m}(\theta,\varphi) e^{i\omega_m t}
				     F_a^{\omega\ell m}(r)
		\, d\omega
							    \text{\punctspace .}
\end{equation}
We also similarly Fourier-transform and spherical-harmonic--expand
the source term (right hand side) $S$ in the
scalar wave equation~\eqref{eqn-theory/scalar-wave-eqn},
\begin{equation}
S(x)
	= \int\limits_\omega
	  \sum_{\ell=0}^\infty
	  \sum_{m=-\ell}^{+\ell} \!
	  Y_{\ell m}(\theta,\varphi) e^{i\omega_m t}
	  S_{\omega\ell m}(r)
		\, d\omega
							    \text{\punctspace .}
\end{equation}
\end{subequations}
%%\end{widetext}

The rest of the mode-sum regularization then goes through unchanged
for each frequency~$\omega$, and each individual scalar-field
mode~$\phi^{\omega\ell m}(r)$ now satisfies the \emph{ordinary}
differential equation
%%\begin{widetext}
\begin{equation}
\frac{d^2 \phi^{\omega\ell m}}{dr^2}
+ H(r) \frac{d\phi^{\ell m\omega}}{dr}
+ V_{\omega\ell}(r) \phi^{\omega\ell m}
	= S_{\omega\ell m}(r)
							    \text{\punctspace ,}
\end{equation}
\end{widetext}
where the coefficients $H(r)$, $V_{\omega l}(r)$, and
$S_{\omega\ell m}(r)$ are again all known analytically.

The frequency-domain approach involves much simpler computations
(ODEs) than the time-domain approach's PDEs.  If the particle orbit
is circular, then only a single frequency is needed, and the integrals
over~$\omega$ are trivial.  On the other hand, if the particle orbit
is noncircular, many frequencies~$\omega$ may be needed to adequately
approximate the integrals over~$\omega$.

In any case, once the individual $F^{\ell m}$ or $F^{\omega\ell m}$
are known, there remains the problem that the overall
$\ell$~sum~\eqref{eqn-theory/mode-sum/F=sum-F-ell} is an infinite
sum.  For computational purposes a finite expression is needed.
The solution to this lies in the known large-$\ell$
behavior~\cite{Detweiler-Messaritaki-Whiting-2003}
of $F_{a,\reg}^\ell := F_a^\ell - [A_a L + B_a + C_a/L]$,
\begin{equation}
F_{a,\reg}^\ell =   \frac{D^{(2)}_a}{L^2}
		  + \frac{D^{(4)}_a}{L^4}
		  + \frac{D^{(6)}_a}{L^6}
		  + \cdots
\qquad
\text{for large $L$}
							    \text{\punctspace ,}
				    \label{eqn-theory/mode-sum/large-ell-series}
\end{equation}
where the coefficients $D^{(k)}_a$ are independent of $L$.  
[Of course, if a term $D^{(k)}_a\!/L^k$ has been added to the
$A_a L + B_a + C_a/L$ inner sum in the mode-sum self-force
formula~\eqref{eqn-theory/mode-sum/F=sum-regularized}, then that term is
absent from the large-$L$ series~\eqref{eqn-theory/mode-sum/large-ell-series}.]
We partition the overall $\ell$~sum~\eqref{eqn-theory/mode-sum/F=sum-F-ell}
into a finite ``numerical'' part and an infinite ``tail'' part,%%%
\footnote{%%%
	 This terminology is perhaps unfortunate: this usage of
	 ``tail'' has no connection at all to the usage of ``tail term''
	 in section~\ref{sect-theory/Detweiler-Whiting}
	 and elsewhere in this article.
	 }%%%
\begin{equation}
F_a = \sum_{\ell = 0}^{\ell_{\max}} F_{a,\reg}^\ell
      +
      \sum_{\ell = \ell_{\max}{+}1}^\infty F_{a,\reg}^\ell
							    \text{\punctspace ,}
\end{equation}
where $\ell_{\max} \sim 15$--$30$ is a numerical parameter.  Then,
once all the $F_{a,\reg}^\ell$ in the numerical part of the sum are
known, we can fit some number (typically~$2$ or~$3$) of terms in the
large-$\ell$ series~\eqref{eqn-theory/mode-sum/large-ell-series}
to the numerically-computed $F_{a,\reg}^\ell$ values, and use the
fitted coefficients $\{D^{(k)}_a\}$ to estimate the tail term.

%%%%%%%%%%%%%%%%%%%%%%%%%%%%%%%%%%%%%%%%

\subsection{The Detweiler-Whiting Decomposition}
\label{sect-theory/Detweiler-Whiting}

Suppose we model the small body as a point particle.
(In this section we ignore the point-particle foundational issues
mentioned in footnote~\ref{footnote-point-particle-problems}.)
Because the particle's own field is singular along the particle
worldline, it's not obvious how to write a meaningful ``perturbation''
theory or formulate equations of motion there.  The solution (within
the framework of modelling the small body as a point particle) is
to somehow regularize the field so that we have finite quantities
to manipulate.

The regularization of the (retarded) field of a point-particle
source near that source is a long-standing problem in mathematical
physics.  Dirac~\cite{Dirac-1938} studied this problem for the
electromagnetic field of a point charge in flat spacetime and
found a decomposition of the field into a singular part which is
spherically symmetric about the charge, and a ``radiative'' part
which is regular at the charge.  
[We might then assume (postulate) that despite being singular, the
spherically-symmetric part exerts no force on the charge.  I discuss
the curved-spacetime generalization of this assumption (postulate) below.]
Dirac's analysis was generalized to curved spacetime by
DeWitt and Brehme~\cite{DeWitt-Brehme-1960} (with a correction by
Hobbs~\cite{Hobbs-1968}).

More recently, Detweiler and Whiting~\cite{Detweiler-Whiting-2003}
found a fully satisfactory curved-spacetime decomposition of a point
particle's field (whether scalar, electromagnetic, or gravitational)
into singular and regular parts.  Here I describe this decomposition
for the gravitational case.  Suppose we have a point mass~$m$,
moving (unaffected by external forces apart from self-force effects)
with 4-velocity $u^a$ along a timelike worldline~$\Gamma \,{=}\, \Gamma(\tau)$
in a ``background'' vacuum spacetime whose typical radius of curvature
in a neighborhood of $\Gamma$ is $\R \gg m$.  [For a LISA EMRI we
might have $m \sim 10 M_\sun$ while $\R \sim 10^6 M_\sun$
(the massive BH's mass).]  

Suppose $\og_{ab}$ is the (vacuum) background metric,
i.e., the spacetime metric in the absence of the particle,
and $\del_a$ is the corresponding covariant derivative.%%%
\footnote{%%%
	 The notation (the presence of the prefix $^{(0)}$ on
	 the background metric, but not on the corresponding
	 covariant derivative operator) is admittedly somewhat
	 inconsistent here.%%%
	 }%%%
{}  Throughout this section we raise and lower indices with the
background metric~$\og_{ab}$.
Let $h_{ab}$ be the metric perturbation produced by the particle,
so that the physical spacetime metric is $g_{ab} = \og_{ab} + h_{ab}$.
We introduce the standard trace-reversed metric
perturbation~$\bar{h}_{ab} = h_{ab} - \thalf \og_{ab} h_c{}^c$
and impose the Lorenz gauge condition $\del_a \bar{h}^{ab} = 0$ on
the metric perturbation.

[The material from this point up to and including the paragraph
containing equation~\eqref{eqn-theory/Detweiler-Whiting-singular-field}
is somewhat more technical than the rest of this article and can
be skipped without creating confusion.]

Up to $\O(m/\R)$~accuracy, the metric perturbation satisfies the
linear (wave) equation
\begin{equation}
\del^c \del_c \bar{h}_{ab} + 2 R^c{}_a{}^d{}_b \bar{h}_{cd} = T_{ab}
							    \text{\punctspace ,}
							 \label{eqn-theory/hbar}
\end{equation}
where $T_{ab}$ is the particle's ($\delta$-function) stress-energy tensor.

\begin{figure}[tb]
\begin{center}
\includegraphics[scale=1.00]{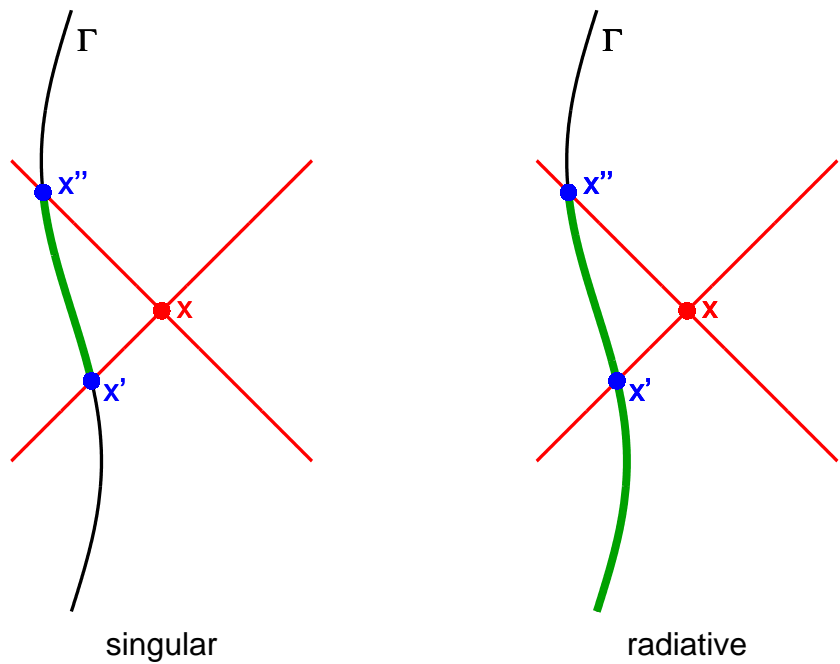}
\end{center}
\vspace*{-5mm}
\caption[Causal Properties of the Detweiler-Whiting Decomposition]
	{%%%
	This figure shows the causal properties of the Detweiler-Whiting
	singular and radiative fields.
	Each subfigure shows (in black) the particle's timelike
	geodesic worldline~$\Gamma$,
	(in red) a nearby ``field'' event $x$
	and its past and future lightcones,
	(in blue) the events $x'$ (past) and $x''$ (future)
	where this lightcone intersects the worldline~$\Gamma$,
	and (in green) the region of the worldline which
	affects the Detweiler-Whiting singular or radiative fields.
	}%%%
\label{fig-theory/Detweiler-Whiting-causality}
\end{figure}

The (linear) perturbation equation~\eqref{eqn-theory/hbar} can
be formally solved by introducing a suitable retarded Green function,
which for events within a normal convex neighborhood of the particle
(i.e., events $x'$ which are linked to the particle event~$x$ by
a \emph{unique} geodesic) can be written in the Hadamard form
\begin{widetext}
\begin{equation}
\G^{ab}{}_{c'd'}(x,x')
	= \U^{ab}{}_{c'd'}(x,x') \, \delta\bigl(\sigma(x,x')\bigr)
	  + \V^{ab}{}_{c'd'}(x,x') \, \theta\bigl(-\sigma(x,x')\bigr)
							    \text{\punctspace ,}
				    \label{eqn-theory/Hadamard-form-of-Green-fn}
\end{equation}
%%\end{widetext}
where $\theta(\cdot)$ is the Heaviside step function,
Synge's world function $\sigma(x,x')$ is half the squared
geodesic distance between the events $x$ and $x'$,%%%
\footnote{%%%
	 As explained very clearly in section~2.1.1 of Poisson's Living
	 Reviews in Relativity article~\cite{Poisson-2004-living-review},
	 Synge's world function $\sigma(x,x')$ has the following properties:
	 \begin{itemize}
	 \item	$\sigma(x,x') < 0$ if and only if the geodesic connecting
		$x'$ to $x$ is timelike, i.e., if and only if $x'$ lies
		within (but not on) $x$'s past light cone.
	 \item	$\sigma(x,x') = 0$ if and only if the geodesic connecting
		$x'$ to $x$ is null, i.e., if and only if $x'$ lies \emph{on}
		$x$'s past light cone.
	 \item	$\sigma(x,x') > 0$ if and only if the geodesic connecting
	 	$x'$ to $x$ is spacelike, i.e., if and only if $x'$ lies
		outside $x$'s past light cone.
	 \end{itemize}
	 }%%%
{} and where $\U^{ab}{}_{c'd'}(x,x')$ and $\V^{ab}{}_{c'd'}(x,x')$
are \emph{smooth} bitensors.%%%
\footnote{%%%
	 For the reader not familiar with bitensors, they are
	 built out of the derivatives $\partial \sigma(x,x')/\partial x$
	 and $\partial \sigma(x,x')/\partial x'$ somewhat analogously
	 to the way that ``ordinary'' tensors are built out of vectors
	 and one-forms.  Section~2.1.2 of Poisson's Living
	 Reviews in Relativity article~\cite{Poisson-2004-living-review}
	 gives a brief and very lucid introduction to bitensors.%%%
	 }%%%
{}  The first term in this decomposition is a singular
``light-cone part'' supported only for $\sigma = 0$, i.e., only
when $x'$ lies on the past light cone of $x$.  The second term
is a smooth ``tail part'' supported only for $\sigma < 0$, i.e.,
only when $x'$ lies within (but not on) the past light cone of $x$.
Notice that the tail part has support throughout the entire past history
of the particle.  Heuristically, this is because metric perturbations
from any time in the particle's past history can scatter off the
background spacetime curvature and then return to the event~$x$.

The resulting formal solution of the perturbation
equation~\eqref{eqn-theory/hbar} is
\begin{equation}
\bar{h}^{ab}(x)
	= \frac{4m}{r} \, \U^{ab}{}_{c'd'}(x,x') u^{c'} u^{d'}
	  + \bar{h}_\tail^{ab}(x)
							    \text{\punctspace ,}
\end{equation}
where $x'$ is the intersection of $x$'s past light cone with the
worldline~$\Gamma$, $u^{a'}$ is the small BH's 4-velocity at this
retarded point $x'$, and the tail term is given explicitly by
\begin{equation}
\bar{h}_\tail^{ab}(x)
	= 4m \int_{-\infty}^{x'}
	     \V^{ab}{}_{c^\dummy d^\dummy} \bigl( x,\Gamma(\tau^\dummy) \bigr)
	     u^{c^\dummy} u^{d^\dummy}
	     \, d\tau^\dummy
				   \label{eqn-theory/tail-term-as-integral-of-V}
\end{equation}
where the $\dummy$~accent (superscript) marks the dummy variable of
integration (integrating over the worldline), and where the integral
is over the particle's entire past history prior to the retarded point~$x'$.
Figure~\ref{fig-theory/worldline-and-lightcone} illustrates the causal
relationships between $\Gamma$, $x$, $x'$, and $x^\dummy$.
Physically, the tail term~\eqref{eqn-theory/tail-term-as-integral-of-V}
models the effect of ``null then scattering then null'' paths from~$x^\dummy$
to~$x$ (shown in blue in figure~\ref{fig-theory/worldline-and-lightcone},
where metric perturbations from the event~$x^\dummy$ scatter off the
background spacetime curvature and then return to the event~$x$.

%%%%%%%%%%%%%%%%%%%%
\begin{figure}[tb]
\begin{center}
\includegraphics[scale=1.00]{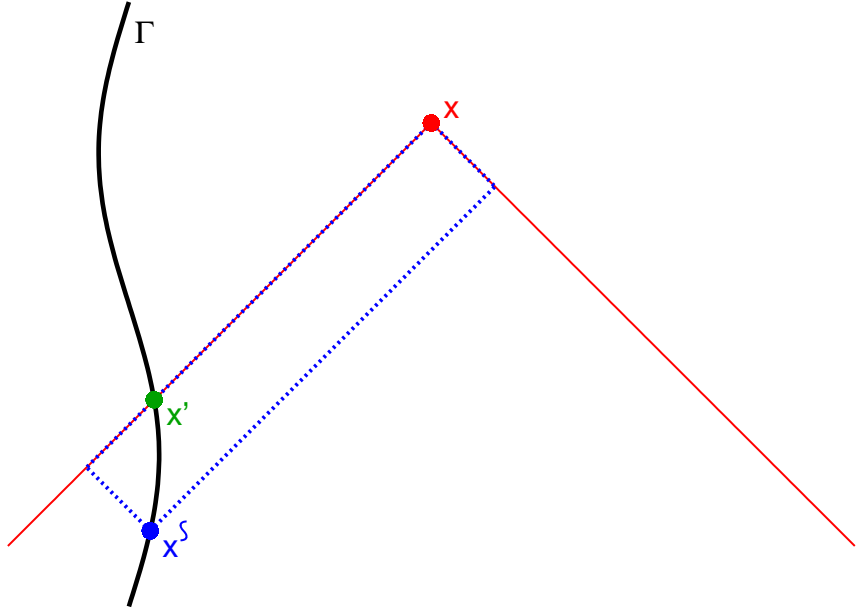}
\end{center}
\vspace*{-5mm}
\caption[Particle Worldline and Field-Point Light Cone]
	{%%%
	This figure shows (in black) the particle's timelike
	geodesic worldline~$\Gamma$,
	(in red) a nearby ``field'' event $x$ and its past lightcone,
	(in green) the event $x'$ where this lightcone intersects
	the worldline~$\Gamma$,
	(in blue) an event $x^\dummy$ on~$\Gamma$ in the past of $x'$,
	and (with blue dashed lines)
	two possible ``null then scattering then null''
	paths from $x^\dummy$ to $x$ (these are modelled by the
	tail term~\eqref{eqn-theory/tail-term-as-integral-of-V}),
	where (heuristically) metric perturbations from the
	event~$x^\dummy$ scatter off the background curvature
	before returning to event~$x$.%%%
	}%%%
\label{fig-theory/worldline-and-lightcone}
\end{figure}
%%%%%%%%%%%%%%%%%%%%

The Detweiler-Whiting ``singular'' part of the metric perturbation
is then (defined to be)
%%\begin{widetext}
\begin{equation}
\bar{h}_S^{ab}(x)
	=   \frac{2m}{r} \, \U^{ab}{}_{c'd'}(x,x') u^{c'} u^{d'}
	  + \frac{2m}{r} \, \U^{ab}{}_{c''d''}(x,x'') u^{c''} u^{d''}
	  - 2m \! \int_{x'}^{x''} \!
		  \V^{ab}{}_{c^\dummy d^\dummy} \bigl(
						x, \Gamma(\tau^\dummy)
						\bigr)
		  u^{c^\dummy} u^{d^\dummy}
		  \, d\tau^\dummy
							    \text{\punctspace ,}
			     \label{eqn-theory/Detweiler-Whiting-singular-field}
\end{equation}
\end{widetext}
where (as shown in figure~\ref{fig-theory/Detweiler-Whiting-causality})
the events~$x'$ and~$x''$ are respectively the intersections of the
past and future light cones of the event~$x$ with the particle
worldline~$\Gamma$.

%%%%%%%%%%%%%%%%%%%%
%%%%%%%%%%%%%%%%%%%%

Detweiler and Whiting showed that the (singular) field~$\bar{h}_S^{ab}$
defined in this manner satisfies the same metric-perturbation
equation~\eqref{eqn-theory/hbar} as the physical (retarded)
perturbation~$\bar{h}^{ab}$, and furthermore that $\bar{h}_S^{ab}$ is
``just as singular'' as $\bar{h}^{ab}$ on the worldline~$\Gamma$.
That is, they showed that the ``radiative'' field
\begin{equation}
\bar{h}_R^{ab}(x)
	= \bar{h}^{ab}(x) - \bar{h}_S^{ab}(x)
\end{equation}
is in fact \emph{smooth} on the worldline~$\Gamma$ (as well as
everywhere else).  Notice also that the radiative field satisfies
the homogeneous form of the metric-perturbation
equation~\eqref{eqn-theory/hbar}.

The Detweiler-Whiting singular and radiative fields
have unusual causal properties, illustrated in
figure~\ref{fig-theory/Detweiler-Whiting-causality}:
the singular field depends on the particle's history only between
the events~$x'$ and~$x''$, while the radiative field depends on
the particle's entire past history up to the advanced event~$x''$.
However, in the limit that the event~$x$ approaches the worldline~$\Gamma$,
the radiative field then depends only on the particle's past history.

The Detweiler-Whiting singular field~$\bar{h}_S^{ab}(x)$ is
spherically symmetric at the particle.  That is, it can be shown%%%
\footnote{%%%
	 See Poisson's \journaltitle{Living Reviews in Relativity}
	 article~\cite{Poisson-2004-living-review} for details.%%%
	 }%%%
{} that if we average the gradient of this field over a 2-sphere of
radius~$\epsilon$ centered on the particle (as seen in the particle's
instantaneous rest frame), then take the limit $\epsilon \to 0$, this
average vanishes.  This motivates the Detweiler-Whiting postulate:
\shortquote{the singular field exerts no force on the particle;
	    the self-force arises entirely from the action of the
	    (regular) radiative field}.
This postulate gives valuable conceptual insight into how the
self force ``works''.  This postulate is also closely linked to the
MiSaTaQuWa equations (I discuss this in the next paragraph and in
section~\ref{sect-theory/MiSaTaQuWa-via-modelling-as-point-particle})
and to puncture-function regularizations and computational schemes for
the self force (I discuss these in section~\ref{sect-theory/puncture-fn}).

Unfortunately, because the field is singular at the particle, the
simple averaging argument described in the previous paragraph doesn't
constitute a rigorous proof of the Detweiler-Whiting postulate.
However, the Detweiler-Whiting postulate is closely linked to the
MiSaTaQuWa equations: if we assume the Detweiler-Whiting postulate,
then the MiSaTaQuWa equations follow almost immediately via the
argument outlined
in section~\ref{sect-theory/MiSaTaQuWa-via-modelling-as-point-particle}.
Because of this close linkage, we can reverse the direction of
logical implication and argue that the validity of the MiSaTaQuWa
equations (which are now well-established via the rigorous derivations
discussed in sections~\ref{sect-theory/MiSaTaQuWa-via-modelling-as-BH}
and~\ref{sect-theory/MiSaTaQuWa-other-derivations}) supports the
correctness of the Detweiler-Whiting postulate.  That is, we can
argue that the Detweiler-Whiting postulate must be valid, since
it is central to a derivation (the one outlined in
section~\ref{sect-theory/MiSaTaQuWa-via-modelling-as-point-particle})
which leads to a correct result (the MiSaTaQuWa equations).
While not a completely rigorous proof, this argument strongly
supports the validity of the Detweiler-Whiting postulate.

Harte~\cite{Harte-2006:EM-self-force,Harte-2008,Harte-2009,Harte-2010}
and Pound~\cite{Pound-PhD,Pound-2010a,Pound-2010b} have recently given
rigorous proofs of the Detweiler-Whiting postulate (somewhat generalized
in some cases).

%%%%%%%%%%%%%%%%%%%%%%%%%%%%%%%%%%%%%%%%

\subsection{Deriving the M\lowercase{i}S\lowercase{a}T\lowercase{a}Q\lowercase{u}W\lowercase{a} Equations via Modelling the Small Body as a Point Particle}
\label{sect-theory/MiSaTaQuWa-via-modelling-as-point-particle}

If we ignore the foundational issues of point particles,%%%
\footnote{%%%
\label{footnote-point-particle-problems}%%%
	 Geroch and Traschen~\cite{Geroch-Traschen-1987} have
	 shown that point particles can \emph{not} consistently
	 be described by metrics with $\delta$-function curvature
	 tensors.  More general Colombeau-algebra methods may
	 be able to resolve this problem~\cite{Steinbauer-Vickers-2006},
	 but the precise meaning of the phrase ``point particle''
	 in general relativity remains a very delicate question.%%%
	 }%%%
{} then the Detweiler-Whiting postulate provides a relatively easy
route to the MiSaTaQuWa equations: the (Detweiler-Whiting) statement
that the self-force arises solely from the action of the (regular)
radiative field is equivalent to the statement that the particle moves
on a geodesic of the background metric perturbed by this radiative
field, i.e., $\og_{ab} + h^R_{ab}$.  The particle's 4-acceleration
is thus given by
\begin{equation}
a^a = \bigl( \og^{ab} + u^a u^b \bigr)
      \bigl( \thalf \del_b h^R_{cd} - \del_d h^R_{bc} \bigr)
      u^c u^d
							    \text{\punctspace .}
					   \label{eqn-theory/point-particle-a^a}
\end{equation}
It's now easy to show that on the particle's worldline~$\Gamma$,
\begin{equation}
\del_c h^R_{ab}
	= -4m \bigl( u_{(a} R_{b)dce} + R_{adbe} u_c \bigr) u^d u^e
	  + \del_c h^\tail_{ab}
							    \text{\punctspace .}
				  \label{eqn-theory/point-particle-del_c-h^R_ab}
\end{equation}
Substituting~\eqref{eqn-theory/point-particle-del_c-h^R_ab}
into~\eqref{eqn-theory/point-particle-a^a} then gives the MiSaTaQuWa
equations
\begin{equation}
a^a = \bigl( \og^{ab} + u^a u^b \bigr)
      \bigl( \thalf h^\tail_{cdb} - h^\tail_{bcd} \bigr)
      u^c u^d
							    \text{\punctspace .}
						   \label{eqn-theory/MiSaTaQuWa}
\end{equation}
where we define $h^\tail_{abc} = \del_c h^\tail_{ab}$.
The corresponding dynamical equations of motion for the particle are
\begin{equation}
u^b \del_b u^a = a^a
						\label{eqn-theory/eqn-of-motion}
							    \text{\punctspace .}
\end{equation}

Notice that the metric $\og_{ab} + h^R_{ab}$ is smooth on the particle
worldline~$\Gamma$.  Moreover, because $\og_{ab}$ is a vacuum solution
of the Einstein equations everywhere and $h^R_{ab}$ satisfies the
homogeneous form of the perturbation equation~\eqref{eqn-theory/hbar},
the metric $\og_{ab} + h^R_{ab}$ is also a vacuum solution of
the Einstein equations everywhere.  This gives what
Poisson~\cite{Poisson-2005-GR17-plenary} describes as
``a compelling interpretation'' to the
condition~\eqref{eqn-theory/point-particle-a^a}:
the particle moves on a geodesic of the vacuum spacetime with
metric~$\og_{ab} + h^R_{ab}$.  Unfortunately, this metric doesn't
coincide with the actual physical spacetime metric~$\og_{ab} + h_{ab}$.

%%%%%%%%%%%%%%%%%%%%%%%%%%%%%%%%%%%%%%%%

\subsection{Deriving the M\lowercase{i}S\lowercase{a}T\lowercase{a}Q\lowercase{u}W\lowercase{a} Equations via Modelling the Small Body as a Black Hole}
\label{sect-theory/MiSaTaQuWa-via-modelling-as-BH}

This point-particle derivation of the MiSaTaQuWa equations is concise,
but depends crucially on the Detweiler-Whiting postulate.  In this
section I outline a different derivation, based on modelling the small
body as a BH.  This derivation doesn't require the assumption of the
Detweiler-Whiting postulate but it (this derivation) is technically
much more involved than the point-particle derivation.  This derivation
is originally due to Mino, Sasaki, and Tanaka~\cite{Mino-Sasaki-Tanaka-1997};
my presentation here is based on that of Poisson's GR17 plenary
lecture~\cite{Poisson-2005-GR17-plenary}.

In this section I adopt the same notation as in the point-particle derivation
(section~\ref{sect-theory/MiSaTaQuWa-via-modelling-as-point-particle})
except that the small body is no longer modelled as a point particle.
Because the small body is (locally) free-falling, its motion is actually
independent of its internal structure (ignoring spin and tidal effects).
This ``effacement of internal structure'' is a fundamental property
of general relativity (\emph{not} shared by most other relativistic
gravity theories) and is discussed in detail in Damour's fascinating
review article in the \booktitle{Three Hundred Years of Gravitation}
volume~\cite{Damour-in-Hawking-Israel-1987}.
In view of this property, we are free to choose the small body's
internal structure for maximum convenience in our analysis; here
we choose it to be a nonrotating (Schwarzschild) BH.

Our analysis will be based on matched asymptotic expansions of the
spacetime metric:
Sufficiently far from the small BH (the ``far zone''), the metric is
that of the background spacetime, perturbed by the presence of the
small BH,
\begin{equation}
g_{ab} = g^\text{background}_{ab} + \O\bigl( m/r \bigr)
							    \text{\punctspace .}
				      \label{eqn-theory/far-zone-g_ab-schematic}
\end{equation}
Sufficiently near to the small BH (the ``near zone''), the metric is
that of the small (Schwarzschild) BH perturbed by the tidal field of
the background spacetime,
\begin{equation}
g_{ab} = g^\text{Schwarzschild}_{ab} + \O\bigl( (r/\R)^2 \bigr)
							    \text{\punctspace .}
				     \label{eqn-theory/near-zone-g_ab-schematic}
\end{equation}
Since $m \ll \R$, there exists an intermediate ``matching zone''
where $m/r \ll 1$ and $r/\R \ll 1$,
and hence both the expansions~\eqref{eqn-theory/far-zone-g_ab-schematic}
and~\eqref{eqn-theory/near-zone-g_ab-schematic} are simultaneously
valid.  In that region these expansions must represent the \emph{same}
vacuum solution of the Einstein equations (modulo gauge choice).
The small BH's motion is then determined by the matching conditions.

[The material from this point up to the start of
section~\ref{sect-theory/MiSaTaQuWa-via-modelling-as-BH/matching}
is somewhat more technical than the rest of this article and can
be skipped without creating confusion.]

We begin by introducing suitable retarded coordinates centered on
the worldline~$\Gamma$:
$v$ is a backwards null coordinate
constant on each ingoing null cone centered on $\Gamma$
(and is equal to proper time on $\Gamma$), and
$r$ is an affine parameter on the cone's null generators.
In this section I use $ijk$ as Penrose abstract indices
ranging over the spatial (non-$v$) coordinates only.
The angular coordinates $\Omega^i = x^i/r$ are constant on each generator
(one can think of $\Omega^i$ as spatial coordinates on a 2-sphere
centered on the small BH).

%%%%%%%%%%%%%%%%%%%%

\subsubsection{The Near-Zone Metric}
\label{sect-theory/MiSaTaQuWa-via-modelling-as-BH/near-zone}

For present purposes it suffices to approximate the spacetime metric
sufficiently near the small BH (the ``near zone'') by that of a
Schwarzschild BH subject to a quadrupole perturbation.%%%
\footnote{%%%
	 This quadrupole term is in general just the
	 leading order in a multipolar expansion.%%%
	 }%%%
{}  In a suitable perturbation $(\tv, \tx{=}\tr\tOmega^i)$ of the
coordinates $(v,x^i{=}r\Omega^i)$, the null-null component of the
near-zone perturbed spacetime metric is
\begin{eqnarray}
g^\near_{\tv\tv}
	& = &	- \left( 1 - \frac{2m}{\tr} \right)
								\nonumber\\
	&   &	- \, \tr^2 \! \left( 1 - \frac{2m}{\tr} \right)^{\!\! 2}
		\E_{ij} \tOmega^i \tOmega^j
		+ \O\bigl( (\tr/\R)^3 \bigr)
							    \text{\punctspace ,}
					       \label{eqn-theory/near-zone-g_uu}
									%%%\\
\end{eqnarray}
where $\E_{ij} = C_{vivj} = \O(1/\R^2)$ are the electric components
of the Weyl tensor $C_{abcd}$; these components measure the tidal
distortion induced by the background spacetime.

%%%%%%%%%%%%%%%%%%%%

\subsubsection{The Far-Zone Metric}
\label{sect-theory/MiSaTaQuWa-via-modelling-as-BH/far-zone}

Sufficiently far from the small BH (i.e., in the ``far zone''), the
null-null component of the background metric is
\begin{equation}
\og_{vv} = - \bigl( 1 + 2ra_i \Omega^i + r^2 \E_{ij} \Omega^i \Omega^j \bigr)
	   + \O\bigl( (r/\R)^3 \bigr)
							    \text{\punctspace ,}
\end{equation}
where $\E_{ij}$ are once again the electric components of the Weyl
tensor, evaluated on~$\Gamma$.

The metric perturbation $\bar{h}_{ab}$ produced by the small BH
satisfies the linear perturbation equation~\eqref{eqn-theory/hbar},
with $T_{ab} = 0$ in the far zone.  Assuming that the far zone lies
within a normal convex neighborhood of the small BH, it can be shown
that the null-null component of the far-zone perturbed spacetime
metric is
\begin{eqnarray}
g^\far_{vv}
	& = &	- \left( 1 - \frac{2m}{r} \right)
								\nonumber\\
	&   &	+ h^\tail_{uu}
								\nonumber\\
	&   &	+ r \bigl(
		    2m \E_{ij} \Omega^i \Omega^j - 2 a_i \Omega^i
		    + h^\tail_{uuu} + h^\tail_{uui} \Omega^i
		    \bigr)
								\nonumber\\
	&   &	- r^2 \E_{ij} \Omega^i \Omega^j
								\nonumber\\
	&   &	+ \O\bigl( (m/\R) (r/\R)^2 \bigr)
		+ \O\bigl( (r/\R)^3 \bigr)
							    \text{\punctspace .}
						\label{eqn-theory/far-zone-g_uu}
									%%%\\
\end{eqnarray}
Differentiating the tail term~\eqref{eqn-theory/tail-term-as-integral-of-V}, we find
that in terms of the original (physical) retarded Green function
$\G^{ab}{}_{c'd'}(x,x'')$, $h^\tail_{abc}$ is given by
\begin{widetext}
\begin{equation}
h^\tail_{abc}
	= 4m \int_{-\infty}^{{x'}^-}
	     \del_c \left(
		    \G_{ab a^\dummy b^\dummy}
		     - \thalf \og_{ab} \G^d{}_{d a^\dummy b^\dummy}
		    \right) (x,x^\dummy)
	     u^{a^\dummy} u^{b^\dummy}
	     \, d\tau^\dummy
							    \text{\punctspace ,}
				   \label{eqn-theory/tail-term-as-integral-of-G}
\end{equation}
\end{widetext}
where the upper integration limit ${x'}^-$ means that the integral
extends over the entire past worldline of the small BH prior to
(but not including) the event~$x'$.  By cutting off the integration
infinitesimally before~$x'$ we include the (regular) tail part of
the Green function and exclude the (singular) light-cone part.
As a result, $h^\tail_{abc}$ is finite (although generally only $C^0$,
i.e., continuous but not differentiable) on the worldline.

%%%%%%%%%%%%%%%%%%%%

\subsubsection{Matching}
\label{sect-theory/MiSaTaQuWa-via-modelling-as-BH/matching}

The coordinate transformation between
the near-zone coordinates~$(\tv,\tx^i{=}\tr\tOmega^i)$ and
the far-zone coordinates~$(v,x^i{=}r\Omega^i)$ can be computed
explicitly (up to sufficient orders in the small-in-the-matching-zone
quantities $m/\R$ and $r/\R$) in terms of $h^\tail_{ab}$, its
integrals and gradients, and $\E_{ij}$.  Using this to transform
the far-zone metric into the near-zone coordinates gives the
null-null metric component
\begin{eqnarray}
g^\far_{\tv\tv}
	& = &	- \left( 1 - \frac{2m}{\tr} \right)
								\nonumber\\
	&   &	- 2\tr \left(
		       a_i - \thalf h^\tail_{uui} + h^\tail_{uiu}
		       \right) \tOmega^i
								\nonumber\\
	&   &	+ (4m\tr - \tr^2) \E_{ij} \tOmega^i \tOmega^j
								\nonumber\\
	&   &	+ \O\bigl( (m/\R) (\tr/\R)^2 \bigr)
		+ \O\bigl( (\tr/\R)^3 \bigr)
							    \text{\punctspace .}
\end{eqnarray}

Requiring that this match the same near-zone
metric component~\eqref{eqn-theory/near-zone-g_uu} up to $\O(m/\R)$
now gives the 3-acceleration of the small BH's worldline as
\begin{equation}
a_i = \thalf h^\tail_{uui} - h^\tail_{uiu}
							    \text{\punctspace ,}
\end{equation}
from which the MiSaTaQuWa equations~\eqref{eqn-theory/MiSaTaQuWa}
follow directly.

Although the full derivation (including all the steps I've omitted in
this brief synopsis) is somewhat lengthy, it can be made quite rigorous,
requiring no unproven assumptions about the physical system.

%%%%%%%%%%%%%%%%%%%%%%%%%%%%%%%%%%%%%%%%

\subsection{Other Derivations of the MiSaTaQuWa Equations}
\label{sect-theory/MiSaTaQuWa-other-derivations}

In this section I briefly mention a number of other derivations of the
MiSaTaQuWa equations.  In the interests of keeping this review both
short and broadly accessible, I won't describe any of these derivations
in detail.

As well as the matched-asymptotic-expansions derivation outlined
in section~\ref{sect-theory/MiSaTaQuWa-via-modelling-as-BH},
Mino, Sasaki, and Tanaka~\cite{Mino-Sasaki-Tanaka-1997} also gave
another derivation based on an extension of the electromagnetic
radiation-reaction analysis of DeWitt and Brehme~\cite{DeWitt-Brehme-1960}.

Quinn and Wald~\cite{Quinn-Wald-1997} took an axiomatic approach,
showing that the electromagnetic self force can be derived by
(i) using a ``comparison axiom'' which relates the electromagnetic
force acting on charged particles with the same charge and
4-acceleration in two possibly-different spacetimes,
and in addition
(ii) assuming that in Minkowski spacetime the half-advanced,
half-retarded electromagnetic field exerts no force on a uniformly
accelerating charged particle.
Quinn and Wald also derived the gravitational self force
(the MiSaTaQuWa equations) using a similar set of axioms.

Building on earlier work by
Harte~\cite{Harte-2006:EM-self-force,Harte-2008,Harte-2009,Harte-2010},
Gralla, Harte, and Wald~\cite{Gralla-Harte-Wald-2009} have recently
provided a rigorous rederivation of the classical (DeWitt-Brehme)
electromagnetic self-force based on taking the limit of a 1-parameter
family of spacetimes corresponding to the small body being ``scaled down''
in charge and mass simultaneously.  (Harte's analysis also includes a
rigorous proof of a generalized form of the Detweiler-Whiting postulate
for the scalar-field and electromagnetic cases, and for the linearized
Einstein equations.)
Gralla and Wald~\cite{Gralla-Wald-2008} have rederived the gravitational
self-force (the MiSaTaQuWa equations) based on a similar technique;
here the small body is ``scaled down'' in size and mass simultaneously.
Both of these derivations are mathematically rigorous and make no
assumptions beyond the existence and appropriate smoothness and limit
properties of the 1-parameter families of spacetimes.

Pound~\cite{Pound-PhD,Pound-2010a,Pound-2010b} has reviewed various
derivations of the MiSaTaQuWa equations (including both the ones I've
outlined here, and others) and developed several new mathematical
techniques for analyzing the self-force problem.  Using these, he
has rederived the MiSaTaQuWa equation in a highly rigorous manner.%%%
\footnote{%%%
	 At the conclusion of his main analysis, Pound writes:
	 ``This concludes what might seem to be the most
	   egregiously lengthy derivation of the self-force
	   yet performed.''.
	 }%%%
{}  His analysis includes a rigorous proof of the (gravitational)
Detweiler-Whiting postulate and also provides many valuable insights
into future directions for the Capra research program; I outline some
of these ``future directions'' in section~\ref{sect-summary/future}.

%%%%%%%%%%%%%%%%%%%%%%%%%%%%%%%%%%%%%%%%

\subsection{Puncture-Function Regularizations}
\label{sect-theory/puncture-fn}

In this section I describe two recently-developed alternate
regularization schemes for self-force computations.  Both schemes
begin by considering a ``residual field'', defined as the difference
between the particle's physical field and a suitably chosen
``puncture function'' which approximates the particle's Detweiler-Whiting
singular field near the particle.  By construction, the residual field
is finite (although of limited differentiability) at the particle
position, and it yields the correct self-force in the force
law~\eqref{eqn-theory/scalar-field-force-law}.  The residual
field satisfies a scalar wave equation similar to the usual
one~\eqref{eqn-theory/scalar-wave-eqn}, except that by construction
the right hand side is now a nonsingular ``effective source'' that
can be calculated analytically.  I describe the puncture function,
the effective source, and the basic outline of how they can be used
to regularize the (singular) field equation in
section~\ref{sect-theory/puncture-fn/basics}.

The puncture function and effective source are constructed to have
certain specified properties near to the particle.  Their behavior
far from the particle can equivalently be described as either
(i) they are undefined far from the particle but the computational
scheme is formulated so as not to make use of them there,
or
(ii) they are defined everywhere but vanish (or are negligibly small)
far from the particle.
Following Wardell and his colleagues~\cite{Wardell-PhD,%%%
Ottewill-Wardell-2008,Ottewill-Wardell-2009,Ottewill-Wardell-2010,%%%
Wardell-Vega-2011:generic-effective-source,%%%
Vega-Wardell-Diener-2011:effective-source-for-self-force},
I use the terminology~(i); note that some other authors use the
terminology~(ii).

Given the puncture function and effective source, Barack and Golbourn's
``$m$-mode'' scheme~\cite{Barack-Golbourn-2007,Barack-Golbourn-Sago-2007,%%%
Dolan-Barack-2011}
does a Fourier decomposition of the resulting equation into
azimuthal~($e^{im\varphi}$) modes, and uses a ``world tube'' technique
to remove any dependence on the puncture function or effective source
far from the particle.  The authors then solve numerically for each
$m$-mode of the residual field (using a time-domain numerical evolution
in $2{+}1$~dimensions for each mode), and compute the final self-force
by summing over all the modes' contributions.  I discuss the Barack-Golbourn
$m$-mode scheme further in section~\ref{sect-theory/puncture-fn/m-mode}.

Vega and Detweiler~\cite{Vega-Detweiler-2008:self-force-regularization,%%%
Vega-etal-2009:self-force-3+1-primer} take a different approach:
Given the puncture function and effective source, they introduce
a smooth ``window function'' to remove any dependence on the puncture
function or effective source far from the particle, then numerically
solve the resulting equation directly in $3{+}1$~dimensions.
I discuss the Vega-Detweiler scheme further in
section~\ref{sect-theory/puncture-fn/Vega-Detweiler}.

For either scheme, there are actually many possible choices for the
puncture function and effective source.  These differ in their tradeoffs
between the difficulty of analytically calculating the puncture function
and effective source, and how accurately the puncture function approximates
the particle's Detweiler-Whiting singular field (and correspondingly,
how small the effective source is and how smooth the puncture function
and effective source are at the particle).  I discuss this further in
section~\ref{sect-theory/puncture-fn/constructing-puncture-fn}.

Throughout this section we consider a point particle of
scalar charge~$q$, moving along a timelike worldline~$\Gamma$ in
(say) Kerr spacetime, whose typical radius of curvature in a
neighborhood of $\Gamma$ is~$\R$.

%%%%%%%%%%%%%%%%%%%%

\subsubsection{The Basic Puncture-Function Scheme}
\label{sect-theory/puncture-fn/basics}

In this section I describe the basic puncture-function regularization in
its simplest form.  This is directly applicable to the Barack-Golbourn
$m$-mode scheme discussed in section~\ref{sect-theory/puncture-fn/m-mode}
but is slightly modified for the Vega-Detweiler scheme discussed in
section~\ref{sect-theory/puncture-fn/Vega-Detweiler}.

We take the scalar field~$\phi$ to satisfy the usual scalar wave
equation~\eqref{eqn-theory/scalar-wave-eqn}.  In general the
Detweiler-Whiting singular field~$\phi^S$ isn't known analytically
but, by careful analysis of the scalar field's singularity structure
near the particle, we can construct approximations to the singular field.
Thus, we define an ``$n$th order puncture function''~$\phi^{S(n)}$ as a
specific approximation -- one that \emph{is} known analytically -- to
the Detweiler-Whiting singular field~$\phi^S$ near the particle,
which satisfies
\begin{equation}
\phi^{S(n)} - \phi^S = \O\bigl( |\lambda|^{n-1} \bigr)
\end{equation}
near the particle, where $\lambda$~is (roughly) the geodesic distance
from the particle (see~\cite{Dolan-Barack-2011} for a precise definition).
Notice that at this point, the puncture function need only be defined
near the particle; in practice, it's usually only defined within at
most a normal convex neighborhood of the particle worldline.  I
discuss the construction of the puncture function further in
section~\ref{sect-theory/puncture-fn/constructing-puncture-fn}.

We define the ``residual'' field
\begin{equation}
\phi^{R(n)} = \phi - \phi^{S(n)}
				   \label{eqn-theory/basics-residual-field-defn}
\end{equation}
near the particle.  The residual field is $C^{n-2}$ at the particle
(and $C^\infty$ elsewhere near the particle) and satisfies the wave
equation
\begin{equation}
\boxop \phi^{R(n)} = S^{\eff(n)}
							    \text{\punctspace ,}
				      \label{eqn-theory/residual-field-wave-eqn}
\end{equation}
with the ``effective source'' $S^{\eff(n)}$ given by
\begin{equation}
S^{\eff(n)}
	= - \boxop \phi^{S(n)}
	  - 4 \pi q \! \int_{-\infty}^{+\infty} \!
		       \frac{\delta^4\bigl(x^a - \Gamma^a(\tau')\bigr)}
			    {\sqrt{-g}}
		       \, d\tau'
						\label{eqn-theory/effective-src}
							    \text{\punctspace ,}
\end{equation}
where as in section~\ref{sect-theory/mode-sum},
the integral extends over the entire worldline of the particle.
In general the effective source is $\O\bigl( \lambda^{n-3} \bigr)$
at the particle (and $C^\infty$ elsewhere near the particle).  

The subtraction in the effective-source
definition~\eqref{eqn-theory/effective-src}
can't be evaluated numerically (both terms are singular at the particle),
but it can be evaluated analytically using a (lengthy) series-expansion
analysis of the field's singularity structure.  I discuss this further
in section~\ref{sect-theory/puncture-fn/constructing-puncture-fn}.

If $\phi^{S(n)}$ is a sufficiently good approximation to the true
Detweiler-Whiting singular field~$\phi^S$ near the particle (i.e.,
if the order~$n$ is large enough), and the Detweiler-Whiting
postulate holds (i.e., the singular field exerts no force on the
particle), then it's easy to see that $q \del \phi^{R(n)}$ at the
particle position gives precisely the desired self-force acting
on the particle.  Thus (apart from the difficulties outlined in the
next two paragraphs) the self-force can be calculated by analytically
calculating the effective source, then numerically solving the wave
equations~\eqref{eqn-theory/residual-field-wave-eqn}, and then
finally taking the gradient of $\phi^{R(n)}$ at the particle
position.

Accurately solving the wave
equation~\eqref{eqn-theory/residual-field-wave-eqn} is made more
difficult by the limited differentiability of the effective source
and residual field at the particle.  With standard finite differencing
methods, this limited differentiability limits the order of
finite-differencing convergence attainable very near the particle.
Current research is exploring a variety of techniques to alleviate
this problem including
ignoring it (i.e., simply accepting the lower order of convergence),%%%
\footnote{%%%
	 It's not clear to me how much of the overall numerical
	 error occurs within a finite-difference-molecule radius
	 of the particle.  If this fraction is small at practical
	 grid resolutions, then a lower order of convergence
	 at those few grid points might have only a minor impact
	 on the overall numerical accuracy.%%%
	 }%%%
{} modifying the finite differencing scheme near the particle,
and using finite-element or domain-decomposition pseudospectral
methods that naturally accommodate well-localized non-differentiability
in the fields~\cite{Sopuerta-etal-2006,Sopuerta-Laguna-2006,%%%
Canizares-Sopuerta-2009a,Canizares-Sopuerta-2009b,%%%
Vega-etal-2009:self-force-3+1-primer}.

Another difficulty with puncture-function regularization schemes is that
in general the puncture function and effective source are only defined
within at most a normal convex neighborhood of the particle whereas
the physically appropriate boundary conditions for the wave
equation~\eqref{eqn-theory/residual-field-wave-eqn}
are applied (to the physical field $\phi$) at infinity.
The Barack-Golbourn $m$-mode scheme and the Vega-Detweiler scheme
take very different approaches to resolving this difficulty;
I discuss these in (respectively)
sections~\ref{sect-theory/puncture-fn/m-mode}
and~\ref{sect-theory/puncture-fn/Vega-Detweiler}
below.

%%%%%%%%%%%%%%%%%%%%

\subsubsection{The Barack-Golbourn $m$-mode Scheme}
\label{sect-theory/puncture-fn/m-mode}

The Barack-Golbourn $m$-mode scheme for self-force
computation~\cite{Barack-Golbourn-2007,Barack-Golbourn-Sago-2007,%%%
Dolan-Barack-2011} begins by defining the puncture function~$\phi^{S(n)}$
and effective source~$S^{\eff(n)}$ exactly as just described
(section~\ref{sect-theory/puncture-fn/basics}).  The authors
then decompose the residual field~$\phi^{R(n)}$ and
effective source~$S^{\eff(n)}$ into Fourier series in the
azimuthal ($\varphi$) direction,
\begin{subequations}
\begin{eqnarray}
\phi^{R(n)}(x)
	& = &	\sum_{m=-\infty}^\infty \!
		\phi^{R(n)}_m(t,r,\theta) e^{im\varphi}
									\\
S^{\eff(n)}(x)
	& = &	\sum_{m=-\infty}^\infty \!
		S^{\eff(n)}_m(t,r,\theta) e^{im\varphi}
									\\
\noalign{\noindent
Away from the particle, the physical scalar field~$\phi$ is smooth
and can be similarly decomposed,}
\phi(x)
	& = &	\sum_{m=-\infty}^\infty \!
		\phi_m(t,r,\theta) e^{im\varphi}
							    \text{\punctspace .}
									%%%\\
\end{eqnarray}
\end{subequations}
Since we're working on a Kerr background, the
wave equation~\eqref{eqn-theory/residual-field-wave-eqn} now separates,
so that each (complex) residual-field $m$-mode $\phi^{R(n)}_m(t,r,\theta)$
satisfies a modified wave equation
\begin{equation}
\boxop_m \phi^{R(n)}_m = S^{\eff(n)}_m
			       \label{eqn-theory/m-mode/residual-field-wave-eqn}
\end{equation}
in $2{+}1$~dimensions, where the operator $\boxop_m$ is easily
derived analytically and where the effective-source $m$~modes are
given explicitly by
\begin{equation}
S^{\eff(n)}_m(t,r,\theta)
	 = \frac{1}{2\pi}
	   \int_{-\pi}^\pi
	  S^{\eff(n)}(t,r,\theta,\varphi') e^{-im\varphi'}
	  \, d\varphi'
							    \text{\punctspace .}
\end{equation}
This integral can be done analytically in some cases, but otherwise
must be done numerically.

There still remains the difficulty that the puncture function and
effective source are only defined near the particle, while the
physical field $\phi$ has has outgoing-wave boundary conditions at
infinity.  To resolve this problem, the authors introduce a worldtube
(whose size is a numerical parameter, and shouldn't be ``too large''
in a sense described below) whose interior contains the particle
worldline~$\Gamma$.  The authors then define a new ``numerical''
field
\begin{equation}
\renewcommand{\arraystretch}{1.333}
\phi^{N(n)}_m
	= \left\{
	  \begin{array}{ll}
	  \phi^{R(n)}_m	& \text{inside the worldtube}		\\
	  \phi_m	& \text{outside the worldtube}      \text{\punctspace ,}
								%%%\\
	  \end{array}
	  \right.
\end{equation}
and solve numerically for this.  The numerical field evidently satisfies
the equations
\begin{equation}
\renewcommand{\arraystretch}{1.333}
\boxop_m \phi^{N(n)}_m
	= \left\{
	  \begin{array}{ll}
	  S^{\eff(n)}_m	& \text{inside the worldtube}		\\
	  0		& \text{outside the worldtube}	    \text{\punctspace .}
								%%%\\
	  \end{array}
	  \right.
\end{equation}
Equivalently, one could say that the authors numerically solve the
equations
\begin{widetext}
\begin{subequations}
				   \label{eqn-theory/m-mode/piecewise-wave-eqns}
\begin{align}
\boxop_m \phi^{R(n)}_m
		& =	S^{\eff(n)}_m
		&&	\text{inside the worldtube}		\\
\boxop_m \phi_m	& =	0
		&&	\text{outside the worldtube}		\\
\phi^{R(n)}_m	& =	\phi_m - \phi^{S(n)}_m
		&&	\text{on the worldtube boundary}    \text{\punctspace .}
				  \label{eqn-theory/m-mode/piecewise-adjust-phi}
									%%%\\
\end{align}
\end{subequations}
\end{widetext}

The authors solve the piecewise
modified wave equation~\eqref{eqn-theory/m-mode/residual-field-wave-eqn}
numerically for each $m$ using a standard time-domain finite-difference
numerical evolution code in $2{+}1$~dimensions.%%%
\footnote{%%%
	 Other numerical methods are of course also possible.%%%
	 }%%%
{}  The authors use arbitrary initial data on a large domain, in the
same manner discussed in section~\ref{sect-Barack-Sago/numerical-soln}
below.

[In a finite-difference numerical code, the piecewise aspect of the
equations~\eqref{eqn-theory/m-mode/piecewise-wave-eqns} is trivial
to implement~\cite{Barack-Golbourn-2007,Dolan-Barack-2011}: the code
stores the grid function~$\phi^{N(n)}_m$, and for each finite differencing
operation, checks if the finite difference molecule crosses the worldtube
boundary.  If so, then the code ``adjusts'' the grid function values
being finite differenced (which in this case might well be a temporary
copy of a molecule-sized region of the actual grid function~$\phi^{N(n)}_m$)
as appropriate using~\eqref{eqn-theory/m-mode/piecewise-adjust-phi}.]

With this scheme neither the puncture function nor the effective source
are ever needed more than a short distance (the maximum finite-difference
molecule size) outside the worldtube.  Hence, so long as the worldtube
isn't too large, it's not a problem that the puncture function and
effective source aren't defined far from the particle.
Outside the worldtube,
the piecewise equations~\eqref{eqn-theory/m-mode/piecewise-wave-eqns}
reduce to $\boxop_m \phi_m = 0$, so it's easy to impose the
appropriate outgoing-radiation outer boundary conditions.

Finally, the authors show that the self-force $F_\self^a$ is given by
\begin{equation}
F_\self^a(\tau)
	= q \sum_{m=0}^\infty \!
	    \Biggl.
	    \left( \del^a \tilde{\phi}^{R(n)}_m \right)
	    \Biggr|_{\Gamma(\tau)}
							    \text{\punctspace ,}
					\label{eqn-theory/m-mode/self-force-eqn}
\end{equation}
where the gradient is evaluated at the particle, and where the
real fields~$\tilde{\phi}^{R(n)}_m$ are defined by
\begin{equation}
\renewcommand{\arraystretch}{1.333}
\tilde{\phi}^{R(n)}_m
	= \left\{
	  \begin{array}{ll}
	  \phi^{R(n)}_m		& \text{if $m = 0$}		\\
	  2 \Realpart\left( \phi^{R(n)}_m e^{im\varphi} \right)
				& \text{if $m > 0$}
							    \text{\punctspace .}
								%%%\\
	  \end{array}
	  \right.
\end{equation}

The infinite sum over~$m$ in the
self-force law~\eqref{eqn-theory/m-mode/self-force-eqn} can be
approximated with a finite computation using a tail-fitting
procedure analogous to that described in
section~\ref{sect-theory/mode-sum}.

The Barack-Golbourn $m$-mode scheme provides a practical and efficient
route to self-force computations for a variety of physical systems.
It is currently the basis for a number of such calculations.  Where
the original mode-sum scheme described in section~\ref{sect-theory/mode-sum}
reduced the self-force problem to the numerical solution of a 2-dimensional
set of PDEs in $1{+}1$~dimensions,%%%
\footnote{%%%
	 I'm describing the time-domain case here; somewhat
	 similar arguments would also apply to a frequency-domain
	 solution.%%%
	 }%%%
{} the $m$-mode scheme reduces the self-force problem to the solution
of a 1-dimensional set of PDEs in $2{+}1$~dimensions.  Both schemes
have the major advantage that the problem-domain size, grid resolution,
and/or other numerical parameters can be varied from one PDE to another.
This greatly increases the efficiency of the numerical solutions.

%%%%%%%%%%%%%%%%%%%%

\subsubsection{The Vega-Detweiler Scheme}
\label{sect-theory/puncture-fn/Vega-Detweiler}

The Vega-Detweiler scheme for self-force
computation~\cite{Vega-Detweiler-2008:self-force-regularization,%%%
Vega-etal-2009:self-force-3+1-primer} takes a somewhat different
approach: it begins by defining the puncture function $\phi^{S(n)}$
and effective source $S^{\eff(n)}$ exactly as described in
section~\ref{sect-theory/puncture-fn/basics}.
The authors then introduce a real $C^\infty$ ``window function''
$W$ chosen (in a manner described further below) such that
\begin{equation}
W = 1 + \O\bigl( (\lambda/\R)^4 \bigr)
\end{equation}
near the particle, and $W \to 0$ ``sufficiently fast'' (i.e., $W$ is
either exactly zero or has decayed to a negligible value) far from
the particle, including in the wave zone and at any event horizon(s)
in the spacetime.

The authors then define the residual field in a manner slightly
different from the definition~\eqref{eqn-theory/basics-residual-field-defn}
of section~\ref{sect-theory/puncture-fn/basics}: using a subscript~$W$
to denote ``windowed'' quantities, the authors define
\begin{equation}
\phi_W^{R(n)} = \phi - W \phi^{S(n)}
							    \text{\punctspace .}
\end{equation}
so that the windowed residual field satisfies the wave equation
\begin{equation}
\boxop \phi_W^{R(n)} = S_W^{\eff(n)}
							    \text{\punctspace ,}
			     \label{eqn-theory/windowed-residual-field-wave-eqn}
\end{equation}
with the windowed effective source given by
\begin{equation}
S_W^{\eff(n)}
	= - \boxop (W\phi^{S(n)} )
	  - 4 \pi q \! \int_{-\infty}^{+\infty} \!
		       \frac{\delta^4\bigl(x^a - \Gamma^a(\tau')\bigr)}
			    {\sqrt{-g}}
		       \, d\tau'
				       \label{eqn-theory/windowed-effective-src}
							    \text{\punctspace ,}
\end{equation}
where once again the integral extends over the entire worldline of
the particle.  By construction, the residual field and effective
source so defined have the same continuity properties at the particle
as described in section~\ref{sect-theory/puncture-fn/basics}.

In the same manner as in section~\ref{sect-theory/puncture-fn/basics},
if the puncture function $\phi^{S(n)}$ is of sufficiently high order
and the Detweiler-Whiting postulate holds, then it's easy to see that
the windowed residual-field gradient~$\del \phi_W^{R(n)}$ at the
particle position gives precisely the desired self-force acting
on the particle.  Thus the self-force can be calculated by analytically
calculating the effective source, then numerically solving the
wave equation~\eqref{eqn-theory/windowed-residual-field-wave-eqn} in
$3{+}1$~dimensions with the
effective source~\eqref{eqn-theory/windowed-effective-src}, then
finally taking the gradient of the windowed residual field~$\phi_W^{R(n)}$
at the particle position.

Since the window function is chosen to approach zero
``sufficiently fast'' far from the particle, it's not a problem for
this scheme that the puncture function and effective source aren't
defined far from the particle.  That is, far from the particle we
have (either exactly or to an excellent approximation) $W = 0$ and
hence $\phi_W^{R(n)} = \phi$ and $S_W^{\eff(n)} = 0$, so the
wave equation~\eqref{eqn-theory/windowed-residual-field-wave-eqn}
becomes simply $\boxop \phi = 0$.  This also makes it easy to impose
the appropriate outgoing-radiation outer boundary conditions on $\phi$.

Like the Barack-Golbourn $m$-mode scheme, the Vega-Detweiler scheme
provides a practical and efficient route to self-force computations
for a variety of physical systems.  The Vega-Detweiler scheme is
designed to reduce the self-force problem to the numerical solution
of a (single) wave equation in $3{+}1$~dimensions.  This type of problem
is quite similar to that solved by many existing $3{+}1$~numerical
relativity codes, so the Vega-Detweiler scheme can often reuse
existing numerical-relativity codes and/or infrastructure.

%%%%%%%%%%%%%%%%%%%%

\subsubsection{Constructing the Puncture Function and Effective Source}
\label{sect-theory/puncture-fn/constructing-puncture-fn}

The key to the success of puncture-function schemes (in either the
Barack-Golbourn or Vega-Detweiler variants) is the construction
of the puncture function~$\phi^{S(n)}$.  This essentially requires
a careful local analysis of the field's singularity structure near
the particle.  This can be done exactly only in very simple cases
(for example, for a static particle in Schwarzschild spacetime).
In more general cases, such an analysis uses lengthy series expansions
and (particularly for higher orders~$n$) is usually done using a
symbolic algebra system.  Once the puncture function is known,
the effective source can then be computed (again symbolically)
directly from the definition~\eqref{eqn-theory/effective-src}.  The
resulting algebraic expressions are very lengthy, so usually the
symbolic algebra system is also used to directly generate C or Fortran
for inclusion in a numerical code.

The difficulty (complexity of the expressions) in computing the
puncture function and effective source in this way rises very rapidly
with the puncture function's order~$n$.  In practice, 4th~order seems
to be both practical and a good compromise between the difficulty
of computing the puncture function and effective source, the expense
of evaluating the resulting (machine-generated C or Fortran) expressions
in a numerical code, and the differentiability (and hence order of
accuracy) of the numerical solution.

Wardell and his colleagues~\cite{Wardell-PhD,%%%
Ottewill-Wardell-2008,Ottewill-Wardell-2009,Ottewill-Wardell-2010,%%%
Wardell-Vega-2011:generic-effective-source,%%%
Vega-Wardell-Diener-2011:effective-source-for-self-force} have
developed efficient software for computing puncture functions and
their corresponding effective sources at (in theory) any order,
and these are now being used in a number of self-force research
projects.  In the interests of brevity I won't try to describe the
details of how these puncture functions are calculated, but
Wardell and Vega~\cite{Wardell-Vega-2011:generic-effective-source}
give a very clear description of this.

%%%%%%%%%%%%%%%%%%%%%%%%%%%%%%%%%%%%%%%%

\subsection{Conservative versus Dissipative Effects}
\label{sect-theory/conservative-vs-dissipative}

The self force can be decomposed into conservative (time-symmetric)
and dissipative (time-antisymmetric) parts.  This decomposition is an
important conceptual tool for understanding the physical meaning of the
self-force.  This decomposition is also important for practical
computations, for reasons described below.

To actually compute the conservative and dissipative parts of the
self force, consider that thus far, we have used solely the
\emph{retarded} scalar field~$\phi^\ret := \phi$ (or metric perturbation
$h^\ret_{ab} := h_{ab}$), and our goal has been to compute the
corresponding \emph{retarded} self-force $F_\ret^a := F_\self^a$.
If we introduce an \emph{advanced} scalar field~$\phi^\adv$
(or metric perturbation $h^\adv_{ab}$) and the corresponding
advanced self-force $F_\adv^a$ (both computed in a manner that's the
time-reversal of that for the corresponding retarded quantities), then
as described in more detail by Dolan and Barack~\cite{Dolan-Barack-2011},
the conservative part of the self-force~$F_\cons^a$ and the
dissipative part~$F_\diss^a$ can easily be computed via
\begin{subequations}
				  \label{eqn-theory/self-force-cons-diss-decomp}
\begin{eqnarray}
F_\cons^a	& = &	\thalf (F_\ret^a + F_\adv^a)
									\\
F_\diss^a	& = &	\thalf (F_\ret^a - F_\adv^a)
				       \label{eqn-theory/dissipative-self-force}
							    \text{\punctspace .}
									%%%\\
\end{eqnarray}
\end{subequations}
This decomposition can also be performed mode-by-mode in a mode-sum or
$m$-mode calculation.  In some cases there are also ways of computing
this decomposition without needing to explicitly compute the advanced
self-force; I describe one such scheme in
section~\ref{sect-Barack-Sago/conservative-vs-dissipative}.

The Detweiler-Whiting singular field is time-symmetric, so it cancels
out in the subtraction~\eqref{eqn-theory/dissipative-self-force} and hence
doesn't affect the dissipative part of the self-force.  This means
that the dissipative part can be computed without regularizing the
singular field, i.e., given a suitable computational scheme, the
dissipative part can be computed much more easily than the conservative
part.  In mode-sum and puncture-function regularization schemes,
the dissipative part of the mode sums also converges much faster
(exponentially instead of polynomially) than the conservative part.

The dissipative part of the self-force directly causes secular
drifts in the small body's orbital energy, angular momentum, and
(for non-equatorial orbits in Kerr spacetime) Carter constant.
Mino~\cite{Mino-2003,Mino-2005:Capra-report,Mino-2005:big-review,%%%
Mino-2006,Mino-2008}
has argued that dissipative self-force alone can be used to calculate
the correct long-term (secular) orbital evolution of an EMRI system:
the conservative part of the self-force appears to cause only
quasi-periodic oscillations in the orbital parameters, not
long-term secular drifts.  This ``adiabatic approximation'' is very
useful and can provide a route to EMRI orbital evolution that's much
simpler and more efficient than the full Capra calculations that are
the main subject of this article.

However, Drasco and Hughes~\cite{Drasco-Hughes-2006} and
Pound and Poisson~\cite{Pound-Poisson-Nickel-2005,Pound-Poisson-2008a}
have found that the adiabatic approximation isn't as accurate as
had previously been thought.  In particular, they have found that
conservative effects also lead to long-term secular changes in the
orbital motion.  Huerta and Gair~\cite{Huerta-Gair-2009} have recently
estimated the magnitude of these latter effects for a quasicircular
EMRI inspiral.  In their approximate model of a LISA EMRI whose GWs
accumulate ${\sim}\, 10^6$~radians of phase in the last year of
inspiral, conservative effects contribute ${\sim}\, 20$~radians
during this time interval.  Conservative effects are likely to
be much larger for eccentric EMRI inspirals.%%%
\footnote{%%%
	 Building on their self-force calculation for arbitrary
	 bound geodesic orbits in Schwarzschild
	 spacetime~\cite{Barack-Sago-2010}
	 (discussed in detail in section~\ref{sect-Barack-Sago}),
	 Barack and Sago~\cite{Barack-Sago-2011} have recently
	 studied conservative self-force effects for eccentric
	 orbits in Schwarzschild spacetime.%%%
	 }%%%
{}  While a small fraction of the total phase, even the
quasicircular-inspiral conservative effects are still large enough
to be easily measurable by LISA (in section~\ref{sect-accuracy/LISA-needs}
I estimate that LISA will be able to detect GW phase differences
as small as ${\sim}\, 10^{-2}$ radians).  Thus conservative effects
must be included to model EMRIs sufficiently accurately for LISA.

Pound and Poisson~\cite{Pound-Poisson-Nickel-2005,Pound-Poisson-2008a}
have also drawn useful distinctions between ``adiabatic'', ``secular'',
and ``radiative'' approximation schemes, which have often been confused
in the past.

%%%%%%%%%%%%%%%%%%%%%%%%%%%%%%%%%%%%%%%%%%%%%%%%%%%%%%%%%%%%%%%%%%%%%%%%%%%%%%%%

\section{Self-Force via the Barack-Ori Mode-Sum Regularization}
\label{sect-Barack-Sago}

In this section I summarize the recent work of
Barack and Sago \cite{Barack-Sago-2010} in which they calculate the
$\O(\mu)$ gravitational self-force on a particle in an arbitrary
(fixed) bound geodesic orbit in Schwarzschild spacetime.

The calculation is done in the Lorenz gauge, decomposing the metric
perturbation due to the particle into tensor spherical harmonics,
solving for the $\ell = 0$ and $\ell = 1$ harmonics via a frequency-domain
method, for the $\ell \ge 2$ harmonics by numerically evolving a
$1{+}1$-dimensional wave equation in the time domain, and then
computing the final self-force using the mode-sum regularization
described in section~\ref{sect-theory/mode-sum}.

This calculation marks a major milestone in the Capra research
program and uses techniques typical of many other self-force
calculations using time-domain integration of the mode-sum--regularized
perturbation equations.  This calculation also illustrates something
of the (high) level of complexity involved in self-force calculations
for astrophysically ``interesting'' physical systems -- the authors
report that (even after many years of preparatory research) it took
over 2~years to develop and debug the techniques and computer code
for this calculation.

In this section I use
$\ell m$ as spherical-harmonic indices, and
$ijk$ as ``tensor component'' indices ranging from 1 to 10
(these indices are always enclosed in parentheses, and index the
individual coordinate components of symmetric rank-2 4-tensors).

%%%%%%%%%%%%%%%%%%%%%%%%%%%%%%%%%%%%%%%%

\subsection{Particle Orbit}
\label{sect-Barack-Sago/particle-orbit}

I take the Schwarzschild line element to be
\begin{subequations}
\begin{eqnarray}
ds^2	& = &	- f \, dt^2
		+ f^{-1} \, dr^2
		+ r^2 (d\theta^2 + \sin^2\theta \, d\varphi^2)
									\\
    	& = &	- f \, du \, dv
		+ r^2 (d\theta^2 + \sin^2\theta \, d\varphi^2)
							    \text{\punctspace ,}
									%%%\\
\end{eqnarray}
\end{subequations}
where
$M$ is the mass of the Schwarzschild spacetime,
$f = 1 - 2M/r$,
$(t,r,\theta,\varphi)$ are the usual Schwarzschild coordinates,
and $(v,u)$ are null coordinates defined by
$v = t + r_*$ and $u = t - r_*$, where
\begin{equation}
r_* = r + 2M \log \left| \frac{r}{2M} - 1 \right|
						\label{eqn-Barack-Sago/r_*-defn}
\end{equation}
is the ``tortise'' radial coordinate.

\begin{figure}[tb]
\begin{center}
\includegraphics[scale=0.85]{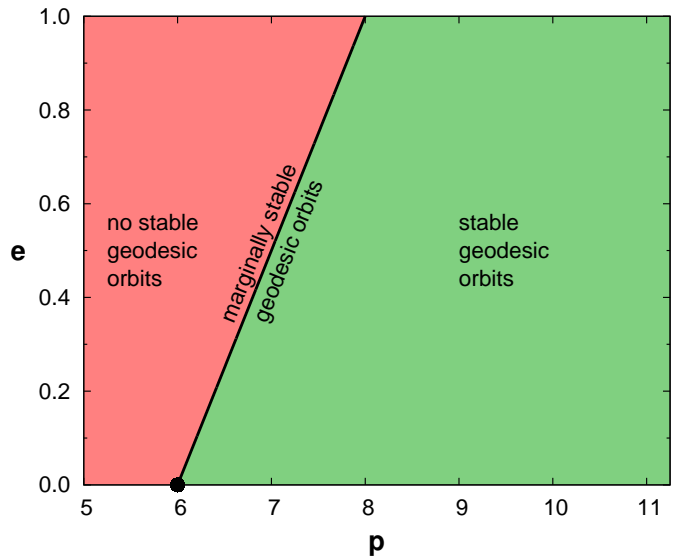}
\end{center}
\caption[Parameter Space of Bound Geodesic Orbits in Schwarzschild Spacetime]
	{%%%
	This figure shows the $(p,e)$ parameter space for bound geodesic
	orbits in Schwarzschild spacetime.  The region $p > 6\,{+}\,2e$
	where stable geodesic orbits are possible is shown in green.
	In the region shown in red ($p < 6\,{+}\,2e$) there are no
	stable geodesic orbits, only unstable ``plunge'' ones.
	The point at $p{=}6$, $e{=}0$ marks the innermost stable
	circular orbit (ISCO).  The line $p = 6\,{+}\,2e$ marks
	the locus of marginally stable orbits, while zoom-whirl orbits
	are those just to the right of this line.
	}%%%
\label{fig-Barack-Sago/Schwarzschild-orbits}
\end{figure}

Without loss of generality the authors take the particle orbit to lie
in the equatorial plane $\theta_p = \tfrac{\pi}{2}$.  The orbit may
be parameterized by its (dimensionless) semi-latus rectum~$p$ and
eccentricity~$e$, defined by
\begin{subequations}
\begin{eqnarray}
p	& = &	\frac{2 r_{\min} r_{\max}}{M (r_{\min} + r_{\max})}	\\
e	& = &	\frac{r_{\max} - r_{\min}}{r_{\max} + r_{\min}}
							    \text{\punctspace ,}
									%%%\\
\end{eqnarray}
\end{subequations}
where $r_{\min}$ and $r_{\max}$ are the minimum and maximum
$r$~coordinate along the orbit.
(For a circular geodesic orbit, $p = r/M$ and $e = 0$.)
Figure~\ref{fig-Barack-Sago/Schwarzschild-orbits} shows the $(p,e)$
corresponding to unstable, marginally stable, and stable orbits.

%%%%%%%%%%%%%%%%%%%%
%%%%%%%%%%%%%%%%%%%%

We normalize $\tau$ (proper time along the particle worldline) to be
zero at a (i.e., at some arbitrary) periastron passage $r = r_{\min}$.
The particle's geodesic motion $x^a = x^a(\tau; p,e)$ can then be
computed by integrating an appropriate set of ODEs~\cite{Sago-2009},
or semi-analytically in terms of elliptic integrals.

%%%%%%%%%%%%%%%%%%%%%%%%%%%%%%%%%%%%%%%%

\subsection{Mode-Sum Regularization}
\label{sect-Barack-Sago/mode-sum}

Using the mode-sum regularization discussed in
section~\ref{sect-theory/mode-sum},
Barack and Ori~\cite{Barack-Ori-2000}, Barack~\cite{Barack-2001},
and Barack~\etal~\cite{Barack-etal-2002}, have shown that the
(Lorenz-gauge) 4-vector gravitational self-force $F^a$ at any
event along the particle's worldline is given by
\begin{equation}
F^a = \sum_{\ell = 0}^\infty F^{a\ell}_\reg
							    \text{\punctspace ,}
				      \label{eqn-Barack-Sago/self-force-sum-ell}
\end{equation}
where the ``regularized self-force mode'' $F^{a\ell}_\reg$ is given by
\begin{equation}
F^{a\ell}_\reg = F^{a\ell}_{\full,\pm}
		   - \Bigl( A^a_\pm (\ell + \thalf) + B^a \Bigr)
							    \text{\punctspace ,}
\end{equation}
where the ``full self-force mode'' $F^{a\ell}_{\full,\pm}$ is
computed for each $\ell$ as described in
section~\ref{sect-Barack-Sago/full-force-modes},
the $\pm$ refers to two different ways of doing this computation
(taking one-sided radial derivatives of the metric perturbation at
the particle from either the outside or inside),
and
$A^a_\pm$ and $B^a$ are ``regularization parameters'' given
semi-analytically in terms of certain elliptic integrals of the
orbit parameters.  Each full self-force mode~$F^{a(\ell)}_{\full,\pm}$
is itself finite at the particle, but their sum diverges whereas
the regularized sum~\eqref{eqn-Barack-Sago/self-force-sum-ell} converges.

The computation of $F^{a(\ell)}_{\full,\pm}$ is based on a
tensor-spherical-harmonic decomposition of the Lorenz-gauge metric
perturbation induced by the particle.  Let
$g_{ab}$ be the background Schwarzschild metric (used to raise and
lower all indices in this section),
$g$ be the determinant of $g_{ab}$,
$\del$ be the corresponding (background) covariant derivative operator,
$h_{ab}$ be the physical (retarded) metric perturbation due to the particle,
$h = h_\mu{}^\mu$ be the trace of $h_{ab}$,
and
$\bar{h}_{ab} = h_{ab} - \thalf g_{ab} h$ be
the trace-reversed metric perturbation.
We take $h_{ab}$ to satisfy the Lorenz gauge condition
\begin{equation}
\del^a \bar{h}_{ab} = 0
							    \text{\punctspace .}
				  \label{eqn-Barack-Sago/Lorenz-gauge-condition}
\end{equation}
Let $u^a$ be the particle's 4-velocity.

To first order in $h_{ab}$, the linearized Einstein equations are
then
\begin{subequations}
				\label{eqn-Barack-Sago/linearized-Einstein-eqns}
\begin{equation}
\del^c \del_c \bar{h}_{ab} + 2 R^c{}_a{}^d{}_b \bar{h}_{cd} = - 16\pi T_{ab}
							    \text{\punctspace ,}
\end{equation}
where
\begin{equation}
T_{ab} = \mu
	 \int_{-\infty}^\infty
	 \frac{u_a u_b \, \delta^{(4)}(x^c - x^c_p)}{\sqrt{-g}}
	 \, d\tau
\end{equation}
\end{subequations}
is the particle's ($\delta$-function) stress-energy tensor.

Due to the symmetry of the Schwarzschild background, the linearized
Einstein equations~\eqref{eqn-Barack-Sago/linearized-Einstein-eqns}
(together with the added constraint-damping terms discussed in
section~\ref{sect-Barack-Sago/numerical-soln}) are
separable into tensorial spherical harmonics via the \German{ansatz}
\begin{equation}
\bar{h}_{ab}
	= \frac{\mu}{r}
	  \sum_{\ell=0}^\infty
	  \sum_{m=-\ell}^{+\ell}
	  \sum_{i=1}^{10}
	  a_\ell^{(i)}
	  \, \bar{h}^{(i)\ell m}(r,t)
	  \, Y_{ab}^{(i)\ell m}(\theta,\varphi; r)
							    \text{\punctspace ,}
\end{equation}
and similarly for $T_{ab}$.

The resulting separated equations take the form of the coupled linear wave
equations
\begin{widetext}
\begin{equation}
\boxop \bar{h}^{(i)\ell m}
+ \sum_{j,a} \N^{(i)\ell}_{(j)a} \partial_a \bar{h}^{(j)\ell m}
+ \sum_j \M^{(i)\ell}_{(j)} \bar{h}^{(j)\ell m}
	= S^{(i)\ell m} \, \delta(r - r_p)
							    \text{\punctspace ,}
					       \label{eqn-Barack-Sago/wave-eqns}
\end{equation}
\end{widetext}
where $\boxop$ is the 2-D scalar wave operator on the Schwarzschild
background, and where
$\N^{(i)\ell}_{(j)a}$, $\M^{(i)\ell}_{(j)}$, and $S^{(i)\ell m}$
are given analytically as known functions of the indices $(i)$
and/or $(j)$, position, $u^a$, and $\ell$ and $m$.%%%
\footnote{%%%
	 The reader is warned that my notation here differs from that
	 of Barack and Sago: I make all derivatives explicit in
	 the wave equations~\eqref{eqn-Barack-Sago/wave-eqns},
	 so that $N$ and $M$ are algebraic \emph{coefficients}, whereas
	 Barack and Sago use $\M$ to denote a single set of 1st-order
	 \emph{differential operators} which contains both the
	 1st~derivative and 0th~derivative terms in the
	 wave equations~\eqref{eqn-Barack-Sago/wave-eqns}.%%%
	 }%%%

The authors solve the wave equations~\eqref{eqn-Barack-Sago/wave-eqns}
numerically to obtain the Lorenz-gauge metric perturbation modes
$\bar{h}^{(i)\ell m}$ and their gradients along the particle worldline.
I describe this numerical solution in
section~\ref{sect-Barack-Sago/numerical-soln}.

%%%%%%%%%%%%%%%%%%%%%%%%%%%%%%%%%%%%%%%%

\subsection{The Full Force Modes}
\label{sect-Barack-Sago/full-force-modes}

Given the Lorenz-gauge metric perturbation modes $\bar{h}^{(i)\ell m}$
in a neighborhood of the particle worldline, the authors next compute
a set of coefficients $f_{(k)\pm}^{a\ell m}$ defined along the particle
worldline in terms of (analytically-known) linear combinations of
$\bar{h}^{(i)\ell m}$, $\partial_{r\pm} \bar{h}^{(i)\ell m}$, and
$\partial_{t\pm} \bar{h}^{(i)\ell m}$, where the $\pm$ corresponds
to taking one-sided derivatives from the outside or inside of the
particle orbit respectively.  (Due to the $\delta$-function
source term in the wave equation~\eqref{eqn-Barack-Sago/wave-eqns},
$\bar{h}^{(i)\ell m}$ is typically $C^0$ near the particle, i.e.,
$\bar{h}^{(i)\ell m}$ is continuous at the particle worldline but
its 1st~derivatives have a jump discontinuity there.)

Taking into account the \emph{tensor} spherical harmonic expansion
of $\bar{h}_{ab}$ and $T_{ab}$ (as compared to the \emph{scalar} spherical
harmonic expansion implicit in the definition of the $f_{(k)\pm}^{a\ell m}$),
the authors then compute
\begin{equation}
F^{a(\ell)}_{\full,\pm}
	= \frac{\mu^2}{r_p^2}
	  \sum_{m=-\ell}^{+\ell}
	  \left(
	  \sum_{p=-3}^{+3} \F_{p\pm}^{a (\ell+p)m}
	  \right)
	  Y^{\ell m}(\theta_p, \phi_p)
							    \text{\punctspace ,}
					     \label{eqn-Barack-Sago/F-full-defn}
\end{equation}
where each $\F_{p\pm}^{a\ell m}$ is a certain (analytically-known)
linear combination of the $f_{(k)\pm}^{a\ell m}$ with the same $\ell$
and $m$.  Because of the definition~\eqref{eqn-Barack-Sago/F-full-defn}
-- and more generally because of the decomposition of tensor spherical
harmonics into scalar spherical harmonics -- a given full force mode
$F^{a(\ell)}_{\full,\pm}$ depends on the Lorenz-gauge metric perturbation
modes $\bar{h}^{(i)\ell' m}$ for $\ell{-}3 \le \ell' \le \ell{+}3$.

%%%%%%%%%%%%%%%%%%%%%%%%%%%%%%%%%%%%%%%%

\subsection{\boldmath{Numerical Solution of the Wave Equations~\eqref{eqn-Barack-Sago/wave-eqns}}}
\label{sect-Barack-Sago/numerical-soln}

For each $(\ell,m)$ the authors solve the 10~coupled
wave equations~\eqref{eqn-Barack-Sago/wave-eqns} numerically for the
10~$\bar{h}^{(i)\ell m}$ fields, using 4th~order finite differencing
on a uniform characteristic (double-null) $(v,u)$ grid.

The authors' ``diamond integral'' finite differencing scheme is adapted
from those of~\cite{Lousto-2005,Haas-2007}.  Since the $\delta$-function
source term in the wave equation~\eqref{eqn-Barack-Sago/wave-eqns} is
nonzero only on the particle's worldline, grid cells away from the
worldline have no source-term contribution, allowing a relatively
straightforward finite differencing scheme. The handling of the source
term for those grid cells which are intersected by the particle
worldline -- or where the finite difference molecule is intersected
by the particle worldline -- is much more complicated, particularly
since (for a non-circular orbit) the particle generally crosses grid
cells obliquely, with no particular symmetry.

Several other aspects of the numerical solution are worth of note here:
\begin{itemize}
\item	The authors found that a direct numerical solution of the
	wave equations~\eqref{eqn-Barack-Sago/wave-eqns} was unstable,
	with rapidly growing violations of the Lorenz gauge
	constraint~\eqref{eqn-Barack-Sago/Lorenz-gauge-condition}.
	Following Barack and Lousto~\cite{Barack-Lousto-2005}, the
	authors added constraint-damping terms to the evolution equations
	so as to dynamically damp these gauge violations.
\item	The correct initial data for the
	wave equations~\eqref{eqn-Barack-Sago/wave-eqns}
	isn't known.  Instead, the authors use zero initial data.
	This results in the evolution initially being dominated by
	spurious radiation induced by the imperfect initial data.
	Fortunately, this spurious radiation dies out (radiates away)
	within a few orbital periods, so in a sufficiently long
	evolution its influence eventually becomes negligible.%%%
\footnote{%%%
	 Recently Field, Hesthaven, and Lau~\cite{Field-Hesthaven-Lau-2010}
	 suggested that some effects of the spurious radiation would
	 in fact \emph{not} die out even after long evolutions.
	 However,
Jaramillo, Sopuerta, and Canizares~\cite{Jaramillo-Sopuerta-Canizares-2011}
	 argue that such ``Jost junk solutions'' are an artifact
	 of a particular (inconsistent) implementation of the
	 $\delta$-function source term.  In practice, almost all
	 time-domain mode-sum self-force calculations -- including
	 the Barack-Sago one being presented here -- ignore this
	 issue with no apparent ill effect.
	 Thornburg~\cite{Thornburg-2010:characteristic-AMR,%%%
Thornburg-2011:highly-accurate-self-force}
	 calculated the self-force to ${\ltsim}\, 1$~part per million
	 relative accuracy using a time-domain mode-sum code which
	 ignored the possibility of Jost (junk) solutions,
	 suggesting that the Jost-solution errors, if present,
	 are very small.
	 }%%%
\item	As always for mode-sum schemes, the numerical calculations
	are done independently for each~$\ell$ and~$m$.  This makes
	the calculation trivial to parallelize.
\item	The length of evolution required (or equivalently, given the
	authors' characteristic grid setup, the size of the grid)
	isn't known \Latin{a~priori}.  Rather, the evolution
	must be long enough (the grid must be large enough)
	for the initial-data spurious radiation to have decayed
	to a sufficiently small level and, more generally, for
	the $\bar{h}^{(i)\ell m}$ field configuration to have
	reached an equilibrium.  In practice, the authors monitor
	$\bar{h}^{(i)\ell m}$ and its gradient along the particle
	worldline, and stop the evolution once these become periodic
	(with the particle-orbit period) to within a numerical
	error threshold of ${\sim}\, 10^{-4}$.  If the fields
	don't meet this criterion before the evolution ends,
	the authors increase the size of the grid and rerun
	the evolution.
\end{itemize}

%%%%%%%%%%%%%%%%%%%%%%%%%%%%%%%%%%%%%%%%

\subsection{Monopole and Dipole Modes}
\label{sect-Barack-Sago/monopole-and-dipole}

The authors were unable to obtain stable numerical evolutions of
the wave equations~\eqref{eqn-Barack-Sago/wave-eqns} for $\ell = 0$
or~$\ell = 1$.  Instead, they
(Barack, Ori, and Sago~\cite{Barack-Ori-Sago-2008})
used a frequency-domain method to solve for $\bar{h}^{(i)\ell m}$
in these cases.

Because $\bar{h}^{(i)\ell m}$ is only $C^0$ at the particle worldline
($\bar{h}^{(i)\ell m}$ is continuous at the particle worldline but
its 1st~derivatives have a jump discontinuity there), a naive
frequency-domain method would have very poor convergence due to
Gibbs-phenomenon oscillations.  The authors
(Barack, Ori, and Sago~\cite{Barack-Ori-Sago-2008}) have developed
an elegant solution to this problem, using the \emph{homogeneous}
modes of the wave equations~\eqref{eqn-Barack-Sago/wave-eqns} as a
basis for the numerical solution.  They report that this ``method
of extended homogeneous solutions'' works very well, with the resulting
frequency-domain Fourier sums converging exponentially fast to the
desired $\bar{h}^{(i)\ell m}$.

%%%%%%%%%%%%%%%%%%%%%%%%%%%%%%%%%%%%%%%%

\subsection{Conservative and Dissipative Parts of the Self-Force}
\label{sect-Barack-Sago/conservative-vs-dissipative}

In general, the technique described in
section~\ref{sect-theory/conservative-vs-dissipative} for decomposing
the self-force into conservative and dissipative parts requires explicitly
(numerically) computing the advanced metric perturbation~$h^\adv_{ab}$
as well as the usual retarded metric perturbation~$h^\ret_{ab}$.
This essentially doubles the overall computational effort.

As an alternative, the authors describe another way of computing the
conservative and dissipative parts of the self-force using only the
usual (retarded) self-force, assuming only that the particle orbit is
periodic with a single intrinsic frequency.  (This is the case for
the authors' system of an arbitrary bound geodesic particle orbit
in Schwarzschild spacetime, as well as for some types of orbits in
Kerr spacetime.)  Given this condition, the authors extend an argument
of Hinderer and Flanagan~\cite{Hinderer-Flanagan-2008} to infer that
\begin{subequations}
\begin{eqnarray}
F_\adv^t      (\tau)	& = &	- F_\ret^t      (-\tau)			\\
F_\adv^r      (\tau)	& = &	+ F_\ret^r      (-\tau)			\\
F_\adv^\theta (\tau)	& = &	+ F_\ret^\theta (-\tau)			\\
F_\adv^\varphi(\tau)	& = &	- F_\ret^\varphi(-\tau)
							    \text{\punctspace .}
									%%%\\
\end{eqnarray}
\end{subequations}
Assuming that the usual $F_\ret^a$ is known for an entire particle
orbit, the conservative and dissipative parts of the self-force can
then be calculated via~\eqref{eqn-theory/self-force-cons-diss-decomp}.
As noted in section~\ref{sect-theory/conservative-vs-dissipative},
this decomposition can also be performed mode-by-mode.

%%%%%%%%%%%%%%%%%%%%%%%%%%%%%%%%%%%%%%%%

\subsection{\boldmath{The $\ell$ Sum}}
\label{sect-Barack-Sago/ell-sum}

The mode sum~\eqref{eqn-Barack-Sago/self-force-sum-ell} is an
\emph{infinite} sum.  For computational purposes a finite expression
is required.  To this end, the authors partition the mode sum,
rewriting~\eqref{eqn-Barack-Sago/self-force-sum-ell} as
\begin{equation}
F^a = \sum_{\ell = 0}^{\ell_{\max}} F^{a(\ell)}_\reg
      +
      \sum_{\ell = \ell_{\max}{+}1}^\infty F^{a(\ell)}_\reg
							    \text{\punctspace ,}
					       \label{eqn-Barack-Sago/F-ell-max}
\end{equation}
where $\ell_{\max} \sim 15$ is a numerical parameter,
and compute the first term numerically.  (This is the
main computation; recall that it requires numerically solving
the wave equations~\eqref{eqn-Barack-Sago/wave-eqns} for
$0 \le \ell \le \ell_{\max}{+}3$.)

To estimate the second term in the partitioned
mode sum~\eqref{eqn-Barack-Sago/F-ell-max}, the authors consider
the conservative and dissipative parts of the self-force separately.
For the conservative part, the authors make use of the known
large-$\ell$ asymptotic series
\begin{equation}
F^{a(\ell)}_{\cons,\reg}
	=   \frac{D_2^a}{(\ell + \thalf)^2}
	  + \frac{D_4^a}{(\ell + \thalf)^4}
	  + \frac{D_6^a}{(\ell + \thalf)^6}
	  + \cdots
							    \text{\punctspace ,}
			      \label{eqn-Barack-Sago/F-ell-reg-large-ell-series}
\end{equation}
where the coefficients $D_k^a$ don't depend on~$\ell$.
The authors least-squares fit the first two terms in this series
to the numerically computed values of $F^{a(\ell)}_{\cons,\reg}$ for
$\ell_{\min} \le \ell \le \ell_{\max}$, where $\ell_{\min} \sim 10$
is another numerical parameter.  Given the fitted coefficients
$\{D_2^a,D_4^a\}$, the second term of the partitioned
mode sum~\eqref{eqn-Barack-Sago/F-ell-max} can then be estimated
in terms of polygamma functions.

The dissipative part of the mode sum~\eqref{eqn-Barack-Sago/F-ell-max}
converges much faster (in fact, exponentially fast) and is thus much
easier to handle numerically: in practice, $F^{a(\ell)}_{\diss,\reg}$
falls below the numerical error even before $\ell = \ell_{\max}$,
so the second term in the partitioned
mode sum~\eqref{eqn-Barack-Sago/F-ell-max} is negligible.

%%%%%%%%%%%%%%%%%%%%%%%%%%%%%%%%%%%%%%%%

\subsection{Results and Discussion}
\label{sect-Barack-Sago/results-and-discussion}

\begin{figure}[tb]
\vspace*{3mm}
\begin{center}
\includegraphics[scale=1.30]{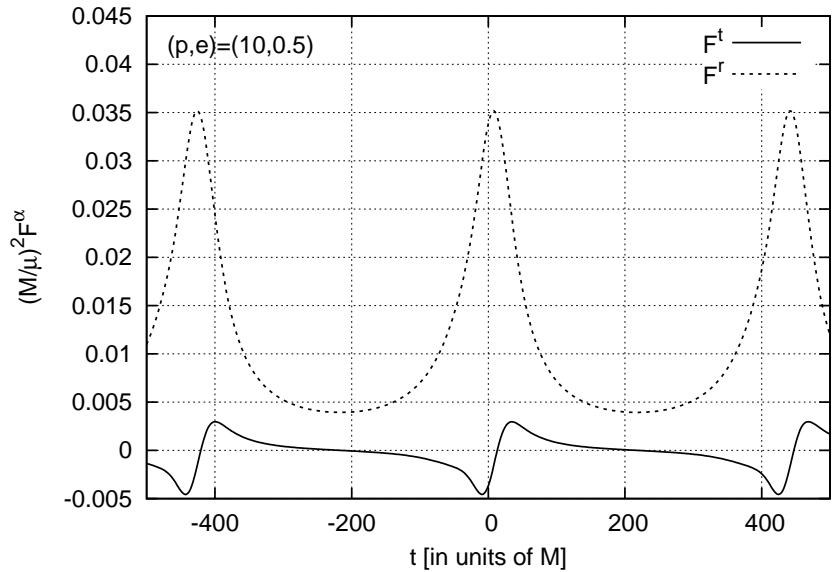}
\end{center}
\vspace*{-3mm}
\caption[Sample Results: Self-Force for various Particle Orbits]
	{%%%
	This figure shows the self-force components 
	$F^r$~(dashed line) and $F^t$~(solid line) as a function
	of scaled Schwarzschild time $t/M$, for a particle orbit
	with $p = 10$ and $e = 0.5$ (so that $r_{\min} = 6\tfrac{2}{3}M$
	and $r_{\max} = 20M$).  The orbital period is $434M$.
	Notice the slight asymmetry of the self-force with respect
	to the orbit (for example, $F^r$ peaks slightly \emph{after}
	the particle's periastron passage at $t/M = 0$).  This is a
	genuine physical effect, not a numerical artifact.
	}%%%
\label{fig-Barack-Sago/sample-results}
\end{figure}

The basic result of the authors' computations is the 4-vector
Lorenz-gauge gravitational self-force $F^a$ as a function of time
along (around) the particle orbit.
Figure~\ref{fig-Barack-Sago/sample-results} shows an example of
these results.

%%%%%%%%%%%%%%%%%%%%
%%%%%%%%%%%%%%%%%%%%

The authors also studied zoom-whirl orbits, finding and analyzing
interesting behavior of the self-force during the whirl phase of the
orbit.  In the interests of brevity, I won't discuss this phenomenon
here.

The authors have also used their code to make the first calculation
of the $\O(\mu)$ self-force corrections to the location and angular
frequency of the ISCO~\cite{Barack-Sago-2009}.  This calculation is
numerically quite delicate, since the ISCO is a singular point in the
$(p,e)$ space of particle orbits
(figure~\ref{fig-Barack-Sago/Schwarzschild-orbits}).
The authors used two different techniques to make the calculation,
and also made a number of other tests to validate the accuracy of
their results.  They found that self-force effects shift the ISCO
inwards, from $r = 6M$ to $\bigl(6 \,{-}\, 3.269 \, \mu\bigr) M$,
and slightly raise its frequency, from $\Omega = 1/(6^{3/2} M)$
to $\bigl(1 \,{+}\, 0.4869 \, \mu/M\bigr) \big/ (6^{3/2} M)$.

These results are of great interest, both as a benchmark of the
current state of the Capra research program and for comparison with
other approaches to modelling EMRI dynamics.  Notably, they can
usefully be compared with post-Newtonian (PN) and effective one-body
(EOB) calculations~\cite{Damour-2010:Schw-grav-self-force-and-EOB,%%%
Barack-Damour-Sago-2010}.  As well as providing valuable tests of
each formalism, this can help to ``calibrate'' various undetermined
coefficients in the PN and EOB expansions.

%%%%%%%%%%%%%%%%%%%%%%%%%%%%%%%%%%%%%%%%%%%%%%%%%%%%%%%%%%%%%%%%%%%%%%%%%%%%%%%%

\section{Accuracy}
\label{sect-accuracy}

In trying to model EMRI orbital dynamics and calculate EMRI GW
templates, it's essential to know what accuracy (in the GW phase)
is needed, and what accuracy is achievable with various approximation
schemes.  In this section I briefly discuss these issues.

%%%%%%%%%%%%%%%%%%%%%%%%%%%%%%%%%%%%%%%%

\subsection{The Accuracy Needed by LISA}
\label{sect-accuracy/LISA-needs}

Matched filtering of the entire years-long LISA data stream would be
impractically expensive for \emph{detecting} EMRIs with hitherto-unknown
parameters~\cite[section~3]{Gair-etal-2004:LISA-EMRI-event-rates}.
However, once EMRIs have been detected by more economical search
algorithms~\cite{Amaro-Seoan-etal-2007:LISA-IMRI-and-EMRI-review,%%%
Porter-2009:LISA-data-analysis-overview}, precision modelling and
matched filtering of the full LISA data set%%%
\footnote{%%%
\label{footnote-transient-resonance-crossings}
	 Flanagan and Hinderer~\cite{Flanagan-Hinderer-2010}
	 have recently found that many LISA EMRI inspirals will
	 include several strong transient resonance crossings.
	 The EMRI osculating-geodesic orbital state vector exiting
	 such a resonance crossing is a very sensitive function
	 of the orbital state vector entering the resonance
	 crossing, with the Jacobian
	 $
	 \partial (\text{post-resonance state})
	 \big/
	 \partial (\text{pre-resonance state})
		\sim \mu^{-1/2} \sim 300
	 $
	 for a canonical $10^6 \,{:}\, 10 M_\sun$ EMRI.  This
	 may limit precision modelling and matched filtering to
	 the intervals between strong resonance crossings, i.e.,
	 to perhaps~$\O(1/3)$ of the full EMRI-inspiral data set.%%%
	 }%%%
{} become practical and even essential to allow detecting and
characterizing weaker sources in the presence of strong EMRIs.

The strongest LISA EMRIs may have signal/noise ratios of
up to $\rho \sim 100$ after
matched filtering~\cite{Amaro-Seoan-etal-2007:LISA-IMRI-and-EMRI-review,%%%
Porter-2009:LISA-data-analysis-overview}, so phase differences on the
order of $1/\rho$~radians should be just detectable in matched filtering.
If we want the maximum possible science return from the LISA mission,
i.e., if we wish to avoid having this science return limited by the
finite accuracy of our GW templates, then these templates should
have GW phase errors of somewhat less than $10$~milliradians.  Any phase
errors larger than this run the risk of significantly increasing the
overall parameter-estimation error budget.  Indeed, if we are lucky
and LISA detects a very strong EMRI with (say) $\rho \sim 300$, the
allowable GW phase errors may be even smaller, perhaps
${\sim}\, 1$--$2$~milliradians.

As a rough approximation, a typical LISA EMRI accumulates
${\sim}\, 10^6$~radians of orbital phase during its last year of
inspiral~\cite{Amaro-Seoan-etal-2007:LISA-IMRI-and-EMRI-review},
so maintaining a phase error of somewhat less than $10$~milliradians
implies a fractional error of somewhat less than $10$~parts per billion%%%
\footnote{%%%
	 I use the North American definition that
	 $\text{billion} = 10^{9}$.%%%
	 }%%%
{} in the instantaneous GW frequency (whose integral gives the cumulative
GW phase), and hence also in the instantaneous EMRI orbital frequency.

It's non-trivial to translate ``required accuracy in the instantaneous
EMRI orbital frequency'' into ``required accuracy in a self-force
calculation'', but we can make a crude estimate using the recent
analyses of Huerta and Gair~\cite[table~1]{Huerta-Gair-2009}.
As discussed in section~\ref{sect-theory/conservative-vs-dissipative},
for a quasicircular inspiral they estimated that $\O(\mu^2)$ self-force
effects (an~$\O(\mu) \sim 10^{-5}$~fraction of the overall self-force)
contribute ${\sim}\, 20$~radians to the cumulative GW phase of our
typical LISA EMRI.  This suggests that a GW phase error tolerance
of ${\ltsim}\, 10$~milliradians corresponds to a fractional accuracy
of roughly
\begin{equation}
\frac{10~\text{milliradians}}{20~\text{radians}} \times 10^{-5}
	= 5 \,{\times}\, 10^{-9}
					   \label{eqn-accuracy/LISA-needs-in-SF}
\end{equation}
in the overall self-force.  This is only a crude estimate, but it
does suggest the general order of magnitude of self-force computation
accuracy needed to match LISA's data quality for strong EMRIs.

%%%%%%%%%%%%%%%%%%%%%%%%%%%%%%%%%%%%%%%%

\subsection{High-Accuracy Capra Computations}
\label{sect-accuracy/high-accuracy}

How accurate is (will) (can) a Capra-based GW template be?
This obviously depends on many factors.  Astrophysically, there are
a variety of possible perturbations to the vacuum--Einstein-equations
(Kerr $+$ compact-object) model used in most Capra calculations to date:
magnetic fields,
accretion disks~\cite{Giampieri-1993,Chakrabarti-1995,Narayan-2000,%%%
Levin-2006,Barausse-Rezzolla-2008},
or even another supermassive BH within a few tenths of a
parsec~\cite{Yunes-Miller-Thornburg-2011}.
It remains an open research problem to incorporate such perturbations
within Capra models.

Within vacuum--Einstein-equations Capra models, there are two major
sources of error in our models of EMRI dynamics and GW emission/propagation:
\begin{itemize}
\item	Our Capra computational schemes are based on approximations
	to the Einstein equations.  For example, the use of 1st~order
	BH perturbation theory implies fractional errors in self-force
	effects of at least $\O(\mu) \sim 10^{-5}$ due to neglected
	2nd-order effects; 2nd~order BH perturbation theory should
	bring these errors down to $\O(\mu^2) \sim 10^{-10}$, subject
	to the long-time-approximation issues discussed in
	section~\ref{sect-accuracy/long-time} below.  No practical
	2nd-order Capra computational schemes exist yet; I discuss
	prospects for their construction in the future in
	section~\ref{sect-summary/future}.
\item	In practice, we numerically solve our equations using
	finite-precision arithmetic, and using discrete approximations
	to the ordinary or partial differential equations (ODEs or PDEs).
	As discussed in section~\ref{sect-theory/mode-sum},
	frequency-domain methods only require integrating ODEs;
	these are potentially very accurate.  For example,
Detweiler, Messaritaki, and Whiting~\cite{Detweiler-Messaritaki-Whiting-2003}
	achieved fractional numerical errors
	${\ltsim}\, 1.5 \,{\times}\, 10^{-8}$
	in a computation of scalar-field self-force acting on
	a scalar particle in a circular orbit in Schwarzschild spacetime,
	and Blanchet \etal{}~\cite{Blanchet-etal-2010:cmp-3PN-with-self-force}
	achieved fractional numerical errors ${\ltsim}\, 10^{-13}$
	in a computation of the gravitational self-force for a
	point mass in a circular orbit in Schwarzschild spacetime.
	However, time-domain methods require numerically solving PDEs,
	and have typical fractional numerical accuracies of at
	best~${\ltsim}\, 10^{-4}$,
	although Thornburg~\cite{Thornburg-2010:characteristic-AMR,%%%
Thornburg-2011:highly-accurate-self-force} achieved fractional
	accuracies~${\ltsim}\, 10^{-6}$ in a time-domain code by
	combining adaptive mesh refinement with extended-precision
	floating-point arithmetic.
\end{itemize}

For very high accuracy \emph{all} error sources need to be small, i.e.,
the EMRI's astrophysical environment must be accurately modelled
\emph{and} the Capra computational scheme must accurately approximate
the Einstein equations (probably using 2nd~order BH perturbation theory
and a long-time approximation of the type described below)
\emph{and} the numerical computations must be very accurate.

%%%%%%%%%%%%%%%%%%%%%%%%%%%%%%%%%%%%%%%%

\subsection{Long-Time Approximation Schemes}
\label{sect-accuracy/long-time}

Most Capra research to date has used 1st~order BH perturbation theory
\emph{and} taken the small body to move on a fixed (timelike) geodesic
worldline in the background Schwarzschild or Kerr spacetime.  Assuming
some computational scheme for the self-force, an obvious improvement
is to use the computed self-force to perturb the particle's worldline
away from being a geodesic, updating a ``deviation vector'' with the
$\O(\mu)$~equations of motion~\eqref{eqn-theory/eqn-of-motion}.
Unfortunately, the worldline is fixed by the 1st-order Bianchi
identity, so it's not obvious that such a scheme can be self-consistent.
As very clearly described by Pound~\cite{Pound-2010a}, a ``gauge relaxation''
technique can be used to allow the particle worldline to vary, but
such a scheme can still be valid for at most a short time: since the
particle's orbit gradually shrinks due to GW emission, the particle's
orbital phase differs from that of a reference geodesic by
${\sim}\, 1$~radian after the ``dephasing time'', which is quite
short -- $\O(\mu^{-1/2})$.  At this point the deviation vector is
large, and the whole approximation scheme breaks down.

A much more sophisticated orbital-evolution scheme is needed to
avoid this problem, i.e., to remain accurate for the (long)
orbital-decay timescales~$\O(\mu^{-1})$.
Hinderer and Flanagan~\cite{Hinderer-Flanagan-2008}
and Pound~\cite{Pound-2010a} discuss various aspects of how such
schemes might be constructed.  The detailed definition and
implementation of such schemes remains a topic for future research.

%%%%%%%%%%%%%%%%%%%%%%%%%%%%%%%%%%%%%%%%%%%%%%%%%%%%%%%%%%%%%%%%%%%%%%%%%%%%%%%%

\section{Summary and Future Prospects}
\label{sect-summary}

%%%%%%%%%%%%%%%%%%%%%%%%%%%%%%%%%%%%%%%%

\subsection{Past Light Cone}
\label{sect-summary/past}

A major area of Capra research has long been the effort to analyze
the singularity structure of the scalar-field, electromagnetic, and
gravitational perturbations induced by point charges/masses.
Much of our current understanding of this structure is based on a
decomposition due to Detweiler and Whiting~\cite{Detweiler-Whiting-2003}.
As discussed in section~\ref{sect-theory/Detweiler-Whiting},
the Detweiler-Whiting decomposition splits the perturbation into
a singular part (which is, in a suitable sense, spherically symmetric
at the particle) and a ``radiative'' part which is finite at the particle.
In this context it's common to assume the Detwiler-Whiting postulate,
which asserts that by virtue of its symmetry the singular field
exerts no force on the particle; self-force effects arise solely
from the particle's interaction with the radiative field.
This postulate has recently been rigorously proved by 
Harte~\cite{Harte-2006:EM-self-force,Harte-2008,Harte-2009,Harte-2010}
and Pound~\cite{Pound-PhD,Pound-2010a,Pound-2010b}.

A variety of lines of reasoning lead to the (same) ``MiSaTaQuWa''
equations of motion for the 1st-order-perturbation-theory self-force
acting on a small body moving in an external field.  The original
derivations of these equations are due to
Mino, Sasaki, and Tanaka~\cite{Mino-Sasaki-Tanaka-1997} and
Quinn and Wald~\cite{Quinn-Wald-1997} (thus the name ``MiSaTaQuWa'').
More recently, the rigorous derivations of
Gralla and Wald~\cite{Gralla-Wald-2008},
Gralla, Harte, and Wald~\cite{Gralla-Harte-Wald-2009},
and Pound~\cite{Pound-PhD,Pound-2010a,Pound-2010b} have helped to
put the MiSaTaQuWa equations on a solid mathematical foundation.

The notion of ``point particle'' has serious foundational
difficulties in a nonlinear field theory such as
general relativity~\cite{Geroch-Traschen-1987}
(see also footnote~\ref{footnote-point-particle-problems}).
However, at least in 1st-order perturbation theory
these difficulties seem to be surmountable.
As Poisson writes~\cite[section~5.5.4]{Poisson-2004-living-review},
	\begin{quote}
	The introduction of a point mass in a nonlinear theory of
	gravitation would appear at first sight to be severely
	misguided.  The lesson learned here is that
	\shortquote{one can in fact get away with it}.
	The derivation of the MiSaTaQuWa equations of motion
	based on the method of matched asymptotic expansions
	does indeed show that results obtained on the basis of a
	point-particle description can be reliable, in spite of
	all their questionable aspects.  This is a remarkable
	observation, and one that carries a lot of convenience:
	It is much easier to implement the point-mass description
	than to perform the matching of two metrics in two
	coordinate systems.
	\end{quote}

The MiSaTaQuWa equations involve a curved-spacetime Green function
which can only rarely be explicitly calculated.  Instead, almost all
practical calculations of self-force effects have returned to the
scalar-field, Maxwell, or Einstein equations (as appropriate), and
regularized the point-particle singularity.

Barack and Ori~\cite{Barack-Ori-2000} 
(see also~\cite{Barack-2000,Barack-2001,Barack-etal-2002,%%%
Barack-Ori-2002,Barack-Lousto-2002,Barack-Ori-2003})
developed the ``mode-sum'' regularization, which provides a practical
route to self-force computations.  As discussed in
section~\ref{sect-theory/mode-sum}, this scheme first decomposes
the field perturbation into spherical harmonics.  Each individual
spherical-harmonic mode can be calculated by numerically solving
a linear wave equation in $1{+}1$~dimensions, or by solving an ODE
if a frequency-domain approach is used.  The self-force is then
obtained by subtracting certain analytically-calculable regularization
parameters from the gradient of each mode's field at the particle,
and finally summing over all modes.

The Barack-Ori mode-sum scheme has been the basis for much further
research as well as having been used for a large number of self-force
calculations in various physical systems.  In section~\ref{sect-Barack-Sago},
I summarize a noteworthy recent self-force calculation using this
scheme, due to Barack and Sago~\cite{Barack-Sago-2010}.  They
calculate the 1st-order-perturbation-theory gravitational self-force
acting on a particle in an arbitrary bound geodesic orbit in Schwarzschild
spacetime, and find and analyze a variety of interesting physical
effects such as the $\O(\mu)$ self-force corrections to the ISCO
position and orbital frequency.  This marks a major milestone in
the Capra research program.

Barack and Golbourn~\cite{Barack-Golbourn-2007}
(see also~\cite{Barack-Golbourn-Sago-2007,Dolan-Barack-2011})
and Vega and Detweiler~\cite{Vega-Detweiler-2008:self-force-regularization}
(see also~\cite{Vega-etal-2009:self-force-3+1-primer}) have recently
developed ``puncture-function'' regularization schemes.  As discussed
in section~\ref{sect-theory/puncture-fn}, these schemes first subtract
a puncture function (a suitable analytically-calculable approximation
to the Detweiler-Whiting singular field) from the physical field
near the particle, leaving a finite ``residual'' field.
The Barack-Golbourn ``$m$-mode'' scheme decomposes the residual field
into a Fourier series in the azimuthal~($\varphi$) direction, and
calculates each azimuthal mode by numerically solving a linear wave
equation in $2{+}1$~dimensions on the Kerr background.  The self-force
is then obtained by summing the field gradient at the particle over
all modes.
The Vega-Detweiler scheme numerically solves the regularized field
equation (a linear wave equation) for the residual field in
$3{+}1$~dimensions, bypassing any mode-sum decomposition.
Both the Barack-Golbourn $m$-mode scheme and the Vega-Detweiler scheme
are now in use for a variety of self-force calculations.

Because of the complexity of self-force calculations, many techniques
have first been developed for scalar-field or electromagnetic particles,
and then extended to the gravitational case.  Historically, self-force
calculations first considered particles moving in Schwarzschild spacetime,
but in recent years a growing number of researchers have considered
particles in Kerr spacetime.

The self force and its effects can usefully be decomposed into
conservative (time-symmetric) and dissipative (time-asymmetric) parts;
the latter are often much easier to calculate.  It was once thought
that accurate EMRI orbital evolutions could be obtained ignoring the
conservative part, but Drasco and Hughes~\cite{Drasco-Hughes-2006} and
Pound and Poisson~\cite{Pound-Poisson-Nickel-2005,Pound-Poisson-2008a}
have found that this isn't the case.  Huerta and Gair~\cite{Huerta-Gair-2009}
have estimated the magnitude of conservative effects as ${\sim}\, 20$~radians
of GW phase (out of a total accumulated GW phase of ${\sim}\, 10^6$~radians)
during the final year of a typical LISA quasicircular EMRI's inspiral;
conservative effects are likely to be much larger for eccentric inspirals.

%%%%%%%%%%%%%%%%%%%%%%%%%%%%%%%%%%%%%%%%

\subsection{Future Light Cone}
\label{sect-summary/future}
Many of the topics discussed in this article remain active areas of
research.  Some of the areas where I expect to see major developments
are:
\begin{itemize}
\item	Further study of the transient resonance crossings recently
	identified by Flanagan and Hinderer~\cite{Flanagan-Hinderer-2010}
	(described in footnote~\ref{footnote-transient-resonance-crossings}).
	These could have a major impact on LISA EMRI data analysis.
\item	Many more self-force calculations for particles orbiting in
	Kerr spacetime.  To this end, several research groups are
	actively working on implementing and extending the
	puncture-function computational schemes described in
	section~\ref{sect-theory/puncture-fn}.
	Warburton and Barack~\cite{Warburton-Barack-2010} have
	also used the ``classic'' Barack-Ori mode-sum scheme
	(discussed in section~\ref{sect-theory/mode-sum})
	for Kerr calculations.
\item	The quantitative comparison of different Capra calculations.
	Sago, Barack, and Detweiler~\cite{Sago-Barack-Detweiler-2008}
	made a very important comparison of this type, showing the
	consistency of a frequency-domain calculation by Detweiler
	and a time-domain calculation by Sago and Barack, despite
	these using different gauges.  Such comparisons serve as
	valuable checks on all the methods and codes involved.
\item	The use of Capra calculations to help determine free
	parameters in post-Newtonian approximation schemes,
	such as the recent work of
	Blanchet \etal{}~\cite{Blanchet-etal-2010:cmp-3PN-with-self-force,%%%
Blanchet-etal-2010:PN-fit-to-Schw-circular-self-force}.
	These comparisons also serve as valuable checks on both
	the Capra and post-Newtonian schemes and computations.
\item	The development of improved long-time approximation schemes,
	initially using simple ``orbit perturbation'' ideas as described
	in section~\ref{sect-accuracy/high-accuracy},
	and later possibly along the lines suggested by
	Hinderer and Flanagan~\cite{Hinderer-Flanagan-2008} and/or
	Pound~\cite{Pound-2010a} (I briefly discuss the need for
	these in section~\ref{sect-accuracy/long-time}).
\item	The development of analyses, and eventually practical
	computational schemes, based on 2nd-order BH perturbation
	theory.%%%
\footnote{%%%
	 Rosenthal~\cite{Rosenthal-2005:2nd-order-scalar-regularization,%%%
Rosenthal-2005:2nd-order-grav-regularization,%%%,
Rosenthal-2006:2nd-order-grav-perturbation,%%%
Rosenthal-2006:2nd-order-grav-self-force}
	 has obtained a formal expression for the 2nd-order
	 self-force, but unfortunately this is in a gauge
	 which is very inconvenient for practical calculations
	 (in this gauge the 1st-order self-force vanishes).%%%
	 }%%%
$^{,}$%%%
\footnote{%%%
	 As well as their use for EMRIs, 2nd-order schemes
	 would be of great value in modelling \emph{intermediate}
	 mass ratio inspirals.%%%
	 }%%%
{}	Pound~\cite{Pound-2010a} has recently reviewed this problem,
	and has suggested at least one possible route to the construction
	of a practical 2nd-order scheme.  The computations required
	are likely to be very complicated (both analytically and
	numerically) but seem to be possible.
\item	The computation of Capra GW templates, and later the
	incorporation of many of the other developments mentioned
	above into the computation of these templates.
\end{itemize}

The result of these and many other developments will (I hope)
eventually be the calculation of highly accurate EMRI orbital
dynamics and GW templates.  The numerical calculations involved in
doing this will almost certainly be very expensive, perhaps comparable
or even larger in magnitude to those for a full numerical-relativity
simulation of a comparable-mass binary BH inspiral/coalescence/ringdown.
Thus it won't be practical to calculate in this way the huge numbers of
GW templates that will be needed for LISA data-analysis template banks.
Rather, moderate numbers of Capra GW templates will be used to
calibrate other (cheaper) approximation schemes (perhaps $n$th-generation
descendents of the ``kludge'' waveforms used today,%%%
\footnote{%%%
	 For an introduction to kludge waveforms see, for example,
	 Barack and Cutler~\cite{Barack-Cutler-2004} or
	 Babak~\etal~\cite{Babak-etal-2007:Kerr-kludge-waveforms,%%%
Babak-etal-2008:Kerr-kludge-waveforms-errata}.%%%
	 }%%%
{} or perhaps new schemes like the effective one-body (EOB) ones
described by Yunes~\cite{Yunes-2009:EOB-EMRI-waveforms-review,%%%
Yunes-etal-2010:EOB-EMRI-waveforms}).  These cheaper schemes
will then be used to generate the actual template banks.

Given the very talented people working on these problems, I predict
that Capra EMRI GW templates meeting the accuracy goal described
in section~\ref{sect-accuracy/high-accuracy} (${\ltsim}\, 10$~milliradians
of phase error over a full million-radian inspiral) will be published
within 10~years of this article's appearance.
I hope many readers of this article will participate in this effort
and that within most of our lifetimes we will see actual LISA data
being filtered with these templates.  There is much work to do.

%%%%%%%%%%%%%%%%%%%%%%%%%%%%%%%%%%%%%%%%%%%%%%%%%%%%%%%%%%%%%%%%%%%%%%%%%%%%%%%%

%%\Section*{Acknowledgements}
\begin{acknowledgments}
I thank Leor Barack for introducing me to the self-force problem,
and Leor Barack, Norichika Sago, Darren Golbourn, Sam Dolan, and
Barry Wardell for many useful conversations.
Leor Barack, Barry Wardell, and Virginia J.~Vitzthum provided
many valuable comments on various drafts of this article.
I thank a referee for many valuable comments
on an earlier version of this article.
\end{acknowledgments}

%%%%%%%%%%%%%%%%%%%%%%%%%%%%%%%%%%%%%%%%%%%%%%%%%%%%%%%%%%%%%%%%%%%%%%%%%%%%%%%%

\bibliographystyle{apsrev4-1}
\bibliography{capra-survey}

%%%%%%%%%%%%%%%%%%%%%%%%%%%%%%%%%%%%%%%%%%%%%%%%%%%%%%%%%%%%%%%%%%%%%%%%%%%%%%%%
%%%%%%%%%%%%%%%%%%%%%%%%%%%%%%%%%%%%%%%%%%%%%%%%%%%%%%%%%%%%%%%%%%%%%%%%%%%%%%%%
%%%%%%%%%%%%%%%%%%%%%%%%%%%%%%%%%%%%%%%%%%%%%%%%%%%%%%%%%%%%%%%%%%%%%%%%%%%%%%%%

\end{document}